\documentclass[letterpaper, 10 pt, conference]{ieeeconf}  

\IEEEoverridecommandlockouts                              

\title{\LARGE \bf
Adaptive Attitude Control for Foldable Quadrotors
} 

\author{Karishma Patnaik and Wenlong Zhang$^{*}$
\thanks{The authors are with School of Manufacturing Systems and Networks, Ira A. Fulton Schools of Engineering, Arizona State University, Mesa, AZ, 85212, USA. Email: {\tt\small $\{$kpatnaik, wenlong.zhang$\}$@asu.edu}.}%
\thanks{$^{*}$Address all correspondence to this author.}%
}
\usepackage{url}
\usepackage{amsmath}
\usepackage{graphicx}
\usepackage{amssymb}
\usepackage{mathtools}
\usepackage{xcolor}
\usepackage{float}
\usepackage{cite}
\usepackage{mwe}
\usepackage[caption=false]{subfig}
\usepackage[pdfencoding=auto, psdextra]{hyperref}
\usepackage{mwe,tikz}\usepackage[percent]{overpic}
\DeclarePairedDelimiter\norm{\lVert}{\rVert}

\newcommand{\rev}[1]{\textcolor{black}{#1}}

\newcommand\copyrighttext{%
	\footnotesize \textcopyright 2023 IEEE. Personal use of this material is permitted.
	Permission from IEEE must be obtained for all other uses, in any current or future 
	media, including reprinting/republishing this material for advertising or promotional 
	purposes, creating new collective works, for resale or redistribution to servers or 
	lists, or reuse of any copyrighted component of this work in other works. 
}
\newcommand\copyrightnotice{%
	\begin{tikzpicture}[remember picture,overlay]
		\node[anchor=south,yshift=10pt] at (current page.south) {\fbox{\parbox{\dimexpr\textwidth-\fboxsep-\fboxrule\relax}{\copyrighttext}}};
	\end{tikzpicture}%
}

\begin{document}

\graphicspath{{figures/}}
\maketitle
\copyrightnotice
\thispagestyle{empty}
\pagestyle{empty}

\begin{abstract}

Recent quadrotors have transcended conventional designs, emphasizing more on foldable and reconfigurable bodies. 
The state of the art still focuses on the mechanical feasibility of such designs with limited discussions on the tracking performance of the vehicle during configuration switching. 
In this article, we first present a common framework to analyse the attitude errors of a folding quadrotor via the theory of switched systems. We then employ this framework to investigate the attitude tracking performance for two case scenarios - one with a conventional geometric controller for precisely known system dynamics; and second, with our proposed morphology-aware adaptive controller that accounts for any modeling uncertainties and disturbances. Finally, we cater to the desired switching requirements from our stability analysis by exploiting the trajectory planner to obtain superior tracking performance while switching. Simulation results are presented that validate the proposed control and planning framework for a foldable quadrotor's flight through a passageway. 

\end{abstract}

\section{Introduction}


  
Foldable quadrotors (FQrs) have created a paradigm shift in the design of multirotor aerial vehicles for flying through small openings and cluttered spaces \cite{PZ21}. While there is ample research demonstrating the mechanical feasibility of the foldable designs \cite{PM+20, PM+21, F+19}, limited literature exists on the analysis of the low-level flight controller and the effects of inflight configuration switching.

The low-level flight control for a FQr is challenging due to the parameter-varying dynamics corresponding to its various configurations. Also, not accounting for any modeling uncertainties, such as inertia or aerodynamics, can further deteriorate the tracking performance. In this context, robust controllers have been explored to obtain the desired tracking performance by considering bounded model uncertainties  \cite{F+21,Z+21,L+19,D+21}. The uncertainty bounds for these systems are generally held constant across the various configurations, and may lead to chattering in the control inputs \cite{SL91}. 

\begin{figure}[t]
    \centering
   \includegraphics[width = 0.48\textwidth]{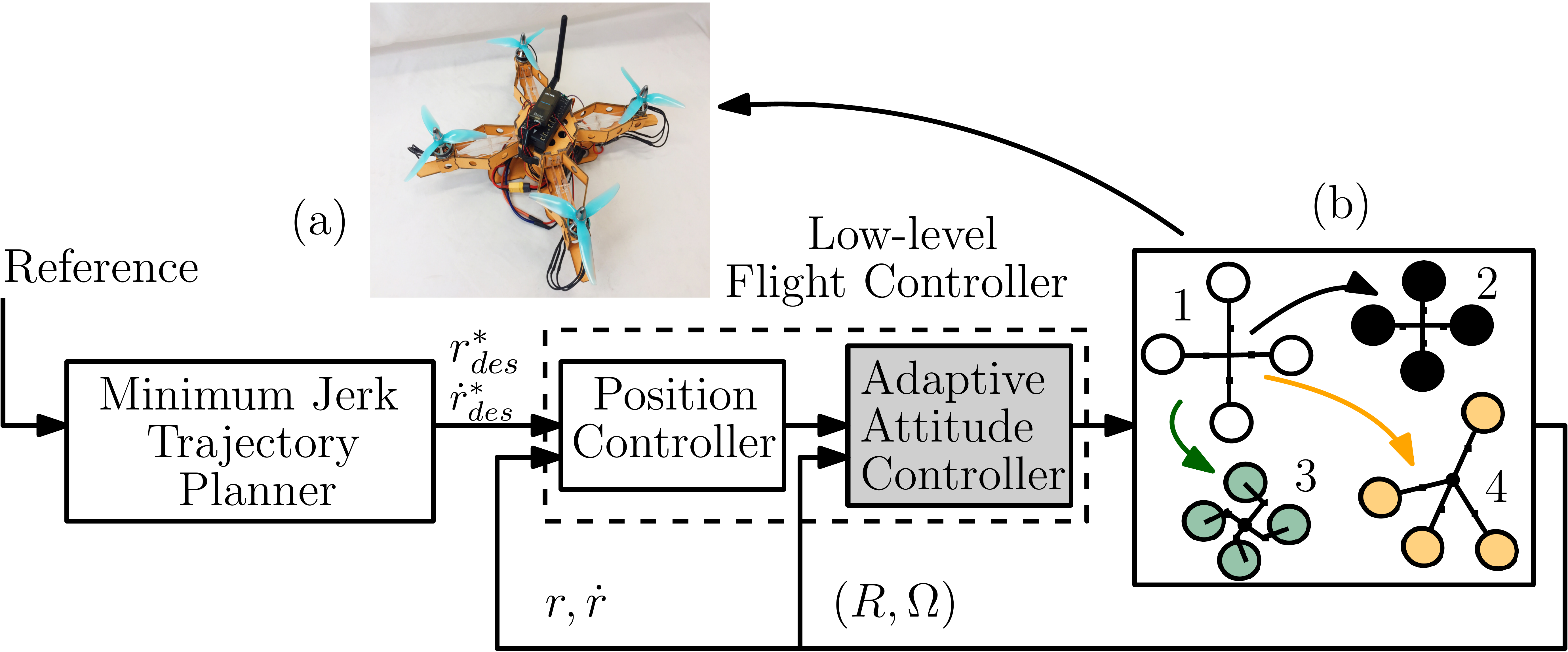}~
    \caption{(a) Example of the foldable quadrotors \cite{Y+19} considered as switching systems in this work. (b) Illustrates a foldable quadrotor switched system consisting of four individual subsystems.} 
    \vspace{-0.3in}
    \label{fig:firstpic}
\end{figure}

Alternatively, adaptive controllers that switch between various operating configurations have also been explored, which fall into the broad category of switched systems (Fig. \ref{fig:firstpic}). For example, researchers have synthesized different LQR controllers for different configurations and the corresponding changes in vehicle dynamics \cite{F+19,BT+21}. 
\rev{Other approaches employing switched model predictive and back stepping controllers have also been developed to address the parameter variation during the change in configuration \cite{PMS21,PN20,DB22}. However, all the aforementioned work assumed precise knowledge of vehicle model. In \cite{BM+22}, the authors proposed an adaptive controller with online parameter estimation, however the rate of change of inertia was assumed negligible, which is not true for switched FQr systems. Furthermore, existing methods fail to address discontinuities encountered during mode switching.} Since the goal of the foldable chassis is to ensure that the vehicle flies through narrow constrained spaces safely, it is important to ensure that this transition-induced disturbance is not significant to cause instability or crashes. Therefore, the switching signal should also be planned, for the transition to occur safely, as a function of vehicle state while adhering to geometric constraints.  

\rev{To the best of the authors' knowledge, this is the first work that introduces a theoretical framework for studying the attitude dynamics of FQrs by modeling them as switched systems. The insights from our analysis are then employed to propose an adaptive controller composed of a parameter estimation law and a robust term, which is duly validated in simulations. We consider three scenarios in our analysis: 1) the simplest case with a precisely-known model, 2) the case with modeling uncertainties in inertia and 3) the  case with external disturbances in addition to unknown inertia.
Furthermore, we propose a coupled control and motion planning framework for FQrs, by augmenting this attitude controller and a PD-type position controller with a \textit{control-aware} minimum-jerk trajectory planner to enforce the stability conditions and guarantee safety during switching.}

The remainder of the paper is organized as follows: 
Section \ref{sec:ps} describes the problem setup with the error definitions in Section \ref{sec:prelims}. Section \ref{sec:control} analyzes the tracking stability for the aforementioned three case scenarios  with the proposed controller while Section \ref{sec:planning} describes the control-aware trajectory generation. Finally, in Section \ref{sec:results}, simulation results are presented that validate the proposed control framework, and Section \ref{sec:conclusion} concludes the letter.

\section{Problem Statement} \label{sec:ps}


Let $x = [{R},~\Omega]^T$ denote the rotation and angular velocity respectively of a foldable quadrotor. Now, consider the following family of systems $\dot{x} = f_p(x)$ corresponding to each configuration shown in Fig. \ref{fig:firstpic}(b) as \rev{\cite{liberzon}}:
\begin{align}\label{eqn:ps}
\centering
        {\dot{R}} &= {R}\hat{\Omega} \nonumber \\ 
        {H}_p\dot{\Omega} &-
        [{H}_p\Omega]_\times \Omega = u + \Delta
\end{align}
with $p \in \mathcal{P}$ where $\mathcal{P} \subseteq  \mathbb{N}$ is the index set and is finite such that $\mathcal{P} = \{ 1,2,...,m\}$. 
To define a switched system generated by the above family, we introduce the \textit{switching signal} as a piece-wise constant function $\sigma : [0,\infty) \rightarrow \mathcal{P}.$ It has a finite number of discontinuities and takes a constant value on every interval between two consecutive switching time instants. The role of $\sigma$ is to specify, at each time instant $t$, the index $\sigma(t) \in \mathcal{P}$ of the active subsystem model from the family (\ref{eqn:ps}) that the FQr currently follows. 
The \textit{hat map} $\hat{\cdot}: \mathbb{R}^3 \xrightarrow{} \mathsf{SO(3)}$ is a symmetric matrix operator defined by the condition that $\hat{x}y = x \times y ~\forall~ x,y \in \mathbb{R}^3$. \rev{The \textit{vee map} $\vee: \mathsf{SO(3)} \xrightarrow{} \mathbb{R}^3$ represents the inverse of the hat map and $[.]_\times$ is the skew symmetric cross product matrix. Further details about the operators are given in the \href{https://arxiv.org/abs/2209.08676}{Appendix}, Section B and C} \cite{appendix}. $\Delta \in \mathbb{R}^3$ represents the disturbances and unmodelled dynamics in the attitude dynamics.


\section{Error Definitions}\label{sec:prelims}
This section describes the definitions of the attitude errors for the tracking problem. The readers are referred to \cite{tLee} for further details.
Consider the error function, $\Phi$, and attitude errors $e_R$ and $e_\Omega$ defined as follows
\begin{equation} \label{eqn:error_definitions1}
\begin{aligned}
    \Phi(R,R_d) &= \frac{1}{2} \text{tr} \Big[ G(I - R_d^TR)\Big] \\
    e_R(R,R_d) &= \frac{1}{2}(GR_d^TR - R^TR_dG)^\vee \\
    e_\Omega(R,\Omega,R_d,\Omega_d) &= \Omega - R^TR_d\Omega_d 
\end{aligned}
\end{equation}
where $G \in \mathbb{R}^{3\times3}$ is given by diag[$g_1, ~ g_2, ~g_3]^T$ for distinct positive constants $g_1, g_2,g_3 \in \mathbb{R}$.
With these definitions, the following statements hold:
\begin{enumerate}
    \item $\Phi$ is locally positive definite about $R = R_d$ 
    \item the left trivialized derivative of $\Phi$ is given by $e_R$
    \item the critical points of $\Phi$ where $e_R = 0$ are \newline 
    $\{R_d\} \bigcup \{R_d \text{ exp}(\pi \hat{s})\} $ for $s \in \{e_1,e_2,e_3\}$ 
    \item the bounds on $\Phi$ are given by
    \begin{equation}\label{eqn:error_function_bounds}
        b_1 \| e_R(R,Rd) \|^2 \leq \Phi(R,Rd) \leq b_2 \| e_R(R,Rd) \|^2
    \end{equation}
\end{enumerate}
The time derivative of the errors are given by
\begin{equation*}
\begin{aligned}
    &\frac{d}{dt}\Phi(R,R_d) = e_R \cdot e_\Omega \\
    \dot{e}_R
    &= \frac{1}{2}(R_d^T R \hat{e}_{\Omega} + \hat{e}_{\Omega} R^T R_d)^\vee \equiv C(R_d^T,R)e_\Omega\\
    \text{with } C(R_d^T,R) &= \frac{1}{2}(\text{tr}[R^T R_dG]I - R^T R_d G)
\end{aligned}
\end{equation*}
It can also be verified that $C(R_d^T,R)$ is bounded by $
    \| C(R_d^T,R) \| \leq \frac{1}{\sqrt{2}}\text{tr}[G]
$. 
Furthermore, 
\begin{equation}
\begin{aligned}
    \dot{e}_\Omega &= \dot{\Omega} + \hat{\Omega} R^T R_d \Omega_d  - R^T R_d \dot{\Omega}_d = \dot{\Omega} - \alpha_D
\end{aligned}
\end{equation}
where $\alpha_D = R^T R_d \dot{\Omega}_d - \hat{\Omega} R^T R_d \Omega_d $ physically represents the angular acceleration term.

\section{Controller Design and Stability Analysis}\label{sec:control}
For this letter, we consider the sub-level set $\mathcal{L} = \{ R_d,R \in \mathsf{SO(3)}|\Phi(R,R_d) < 2\}$ such that the initial attitude error satisfies $\Phi(R(0),R_d(0)) < 2$.
Note that this requires that the initial
attitude error should be less than 180$^o$. Future extensions of this work will analyze complete low-level flight controller stability over the entire $\mathsf{SO(3)}$.  
In this section, we will first provide the methodology for stability analysis for attitude tracking of FQrs modeled as switched systems. We present the conditions for switching such that the overall system retains the tracking performance when the system model and parameter values are precisely known. Next, we propose an adaptive controller which estimates the unknown inertia online and extend the stability analysis with the proposed controller. 

\subsection{Case with the precise model, $\Delta = [0~0~0]^T$}
For this case scenario,  ${H}_p$ is precisely known for each $p^{th}$ subsystem in (\ref{eqn:ps}).
\subsubsection{Attitude tracking of individual subsystems}
The attitude dynamics for an individual subsystem from the switched system of (\ref{eqn:ps}) can be rewritten in the form of ${H_p}{\dot{\Omega}} - {Y}_1 {h}_p  = {u}$ where ${Y}_1 \in \mathbb{R}^{3\times3}$ and ${h}_p = [h_{p_{xx}}~h_{p_{yy}}~h_{p_{zz}}~h_{p_{xy}}~h_{p_{xz}}~h_{p_{zz}}]^T$is the vector encompassing the unique elements of the moment of inertia tensor.

The control moment in this case can be generated according to (\ref{eqn:control_tau_1}) as proposed in \cite{L+10}. 
\begin{align}
        u &= -k_R {e}_R - k_\Omega {e}_\Omega - {Y}{h}_p  \label{eqn:control_tau_1} 
\end{align}
where ${Y} = {Y}_1 - {Y}_2$ with ${H}_p \alpha_d \triangleq {Y}_2 {h}_p$. The exact definitions of $Y_1$ and $Y_2$ are given in \href{https://arxiv.org/abs/2209.08676}{Appendix}, Section D and E respectively.
\subsubsection*{Proposition 1} For positive constants $k_\Omega$ and $k_R$, if the positive constant $c$ is chosen such that 
\begin{equation}
\begin{aligned}\label{eqn:c}
         c_1 < \text{min} &\Bigg \{ \frac{\sqrt{2}k_\Omega}{\text{tr}[G]}, \frac{4\sqrt{2} k_R k_\Omega {(\Lambda_{min}^p})^2}{\sqrt{2} k_\Omega^2 \Lambda_{max}^p+ 4 k_R {(\Lambda_{min}^p)}^2 \text{tr}[G]},\\
         & \sqrt{b_1k_R\Lambda_{min}^p}, \sqrt{b_2k_R \Lambda_{max}^p} \Bigg\}
\end{aligned}
\end{equation}
then the attitude tracking dynamics of the individual subsystems, ($e_R,e_\Omega$), are exponentially stable in the sublevel set $\mathcal{L}$. Moreover, if each subsystem resides in a particular switched state for a minimum dwell-time given by $\tau_d$ in (\ref{eqn:tau_d}), the switched system in (\ref{eqn:ps}) is asymptotically stable in $\mathcal{L}$. Here, $\Lambda_{max}^{(\cdot)}$ and $\Lambda_{min}^{(\cdot)}$ refer to the maximum and minimum eigen values respectively of the quantity $(\cdot)$ and $W_2^p$ is defined as (\ref{eqn:W2}) $\forall~ p \in \mathcal{P}$.
\begin{equation}\label{eqn:tau_d}
    \tau_d > \frac{1}{2(\sum\beta_i)}\text{log}\frac{\prod \Lambda_{max}^{W_2^p} }{\prod \Lambda_{min}^{W_1^p}}, p = 1,2..m \in \mathcal{P}
\end{equation}

\subsubsection*{Proof}
Here we provide a brief sketch of the stability of the attitude tracking errors for the individual subsystem. For full details, the readers are referred to \cite{appendix}.

Consider the individual subsystem's Lyapunov candidate $\forall p = 1,2..m \in \mathcal{P}$ as
\begin{equation}
\begin{aligned}
          \mathcal{V}_p =~& \frac{1}{2}e_\Omega^T {H_p}e_\Omega + k_R\Phi(R,R_d) + c_1 {e}_R \cdot {e}_\Omega 
\end{aligned}
\end{equation}
In the sub-level set $\mathcal{L}$, 
\begin{equation}\label{eqn:Vbounds}
    \Lambda_{min}^{W_1^p} \| z_1 \|^2 \leq \mathcal{V}_p \leq  \Lambda_{max}^{W_2^p} \| z_1 \|^2
\end{equation}
where $z_1= [\| {e}_R \|~ \|{e}_\Omega \|]^T$ and $W_1^p$, $W_2^p \in \mathbb{R}^{2\times2}$ are 
\begin{equation}\label{eqn:W2}
    W_1^p = \frac{1}{2}\begin{bmatrix}
        b_1k_R & -c_1\\
        -c_1 & \Lambda_{p_{min}} \\
    \end{bmatrix},
    W_2^p = \frac{1}{2}\begin{bmatrix}
        b_2k_R & c_1 \\
        c_1 & \Lambda_{p_{max}} \\
    \end{bmatrix}
\end{equation}
We can show that (similar to the proof in \cite{L+10})
\begin{equation}\label{eqn:Vexp}
     \dot{\mathcal{V}}_p \leq -2\beta_p \mathcal{V}_p
\end{equation}
where $\beta_p = \frac{\Lambda_{min}^{W_3^p}}{2\Lambda_{max}^{W_2^p}}$.
Hence the tracking errors are exponentially stable for the individual subsystems. This implies that if $\sigma(t) = p$ for $t \in [t_0,t_0 + \tau_d)$, we have
\begin{equation}\label{eqn:tau_d_exp}
    \mathcal{V}_p (z_1(t_0 + \tau_d)) \leq e^{-2\beta_p\tau_d}\mathcal{V}_p(z_1(t_0))
\end{equation}
\subsubsection{Stability of the overall switched system} 
We will use multiple Lyapunov functions to prove the stability of the switched system. Consider the following Lemma 2:
\subsubsection*{Lemma 2(\cite{liberzon}, pp 41-42)} 
\textit{
Consider a finite family of globally asymptotically stable systems, and let $\mathcal{V}_p$, $p \in \mathcal{P}$ be a family of corresponding radially unbounded Lyapunov functions. Suppose that there exists a family of positive definite continuous functions $\mathcal{W}_p, p \in \mathcal{P}$ with the property that for every pair of switching times ($t_i,t_j)$, $i < j$, such that $\sigma(t_i) = \sigma(t_j)$ and $\sigma(t_k) \neq p$ for $t_i < t_k < t_j$, we have}
\begin{equation}
    \mathcal{V}_p(x(t_j)) - \mathcal{V}_p(x(t_i)) \leq -\mathcal{W}_p (x(t_i)),
\end{equation}
\textit{then the switched system (\ref{eqn:ps}) is globally asymptotically stable.}

\subsubsection*{Proof} Employing (\ref{eqn:Vexp}), we can find a desired lower bound on the \textit{dwell-time}, that corresponds to the amount of time that a system should reside in subsystem $p$ to ensure that the overall tracking errors converge to zero. To elaborate, consider a system when $\mathcal{P} = \{ 1,2\}$ and $\sigma$ takes values of 1 on [$t_0, t_1$) and 2 on  [$t_1, t_2$) such that $t_{i+1} - t_i \geq \tau_d, i = 0,1$. From (\ref{eqn:tau_d_exp}), the minimum dwell-time can be calculated using the theory of the switched systems \cite{liberzon} as (\ref{eqn:tau_d}), which guarantees that the switched system (\ref{eqn:ps}) is asymptotically stable in $\mathcal{L}$ by employing \textit{Lemma 2}.

\subsubsection*{Remark 3} \rev{Since the active reconfigurable quadrotors are designed to avoid collisions while flying through narrow gaps, by strictly adhering to the dwell time obtained in (\ref{eqn:tau_d}) and not allowing for the configuration switching, can be conservative. Hence the trajectory planner is designed to choose the switching signal trajectory, $\sigma(t)$, by accounting for both, the dwell time and also the geometric space constraints, as discussed in Section \ref{sec:planning}}.

\subsection{Case with model uncertainties in  $\textit{{H}}_\textit{p}$ , $\Delta = [0 ~0 ~0]^T$} \label{sec:caseB}
The dwell-time derived in (\ref{eqn:tau_d}) ensures that the switched system is stable when the model (e.g., moment of inertia) is known. However, this is not the case for almost all real-world scenarios. To handle modeling errors, we will estimate the moment of inertia online for each subsystem.

There have been many approaches to estimate the moment of inertia online \cite{tLee}, however only recently researchers have tried to ensure physical consistency of the inertia estimates \cite{L+18,LS21}. For this work, we aim to ensure physical consistency during adaptation of the inertia parameters and hence adopt the methodology presented in \cite{L+18}.

\subsubsection{Attitude tracking for individual subsystems}
For the $p^{th}$ subsystem, let us assume that the control torques are now generated according to 
\begin{align}
        u &= -k_R {e}_R - k_\Omega {e}_\Omega - {Y}{\hat{h}}_p \label{eqn:control_torqueJhat}, \\
    {\dot{\hat{h}}}_p &=
        - (\nabla^2\psi({\hat{h}}_p))^{-1} {Y}^T {e}_A, \\
        {e}_A &= {e}_\Omega + c_2 {e}_R,
        \label{eqn:updaterate_h}
\end{align}
where the inertia parameters are estimated based on the augmented error ${e}_A$. 
Here, $\psi(\cdot)$ is the log-determinant function which ensures that the estimates of the inertia parameters are physically consistent given that the initial guess is also physically consistent. 

\subsubsection*{Assumption 4} The minimum eigen value $\Lambda_{max}^p$ and the maximum eigen values $\Lambda_{min}^p$ of the true inertia matrix ${H}_p$ for the $p^{th}$ subsystem are known. 

\subsubsection*{Proposition 5} Suppose that Assumption 4 holds. For the control generated according to (\ref{eqn:control_torqueJhat})-(\ref{eqn:updaterate_h}), with positive constants $k_\Omega$ and $k_R$ in , if the positive constant $c$ is chosen such that (\ref{eqn:c_hdot}) holds, the attitude tracking errors, ($e_R,e_\Omega$), for the individual subsystems converge to zero asymptotically.
\begin{equation}
\small
\begin{aligned}\label{eqn:c_hdot}
         c_2 < \text{min} &\Bigg \{\sqrt{\frac{2b_1k_R\Lambda_{min}^p}{(\Lambda_{max}^p)^2}}, \frac{\sqrt{2}k_\Omega}{\Lambda_{max}^p \text{tr}[G]}, 
         &\frac{4 k_R k_\Omega}{k_\Omega^2 + \frac{4}{\sqrt{2}} k_R\Lambda_{max}^p \text{tr}[G]}\Bigg\}
\end{aligned}
\end{equation}

\subsubsection*{Proof} 
We will again proceed to first analyze the stability of the individual system and the stability of the switched system. Consider the Lyapunov candidate for individual subsystem as the following
\begin{equation}\label{eqn:lyapunov_p}
\begin{aligned}
          \mathcal{V}_p =~& \frac{1}{2}e_\Omega^T {H_p}e_\Omega + k_R\Phi(R,R_d) + c {e}_R \cdot {H_p}  {e}_\Omega + d_\psi ({h}_p \| {\hat{h}})
\end{aligned}
\end{equation}
where $ d_\psi ({h}_p \| {\hat{{h}}_p})$ is the Bregman divergence operator \cite{L+18}:
\begin{equation*}
\begin{aligned}
    d_\psi ({h}_p \| {\hat{{h}}_p}) = \psi({h}_p) - \psi(\hat{{h}}_p) - ({h}_p - \hat{{h}}_p)^T\nabla \psi(\hat{{h}}_p)
\end{aligned}
\end{equation*}
and the time-derivative of $d_\psi ({h}_p \| {\hat{h}}_p)$ is
\begin{equation}
\begin{aligned}
          \dot{d_\psi }(\cdot) & =
        (\hat{{h}}_p - {h}_p)^T\nabla^2 \psi(\hat{{h}}_p)\dot{\hat{{h}}}_p
\end{aligned}
\end{equation}
As shown in \cite{L+18}, $d_\psi ({h}_p \| {\hat{h}}_p)$ can be taken as an approximation for the geodesic estimation error with the properties required of a desired Lyapunov candidate. Also, from (\ref{eqn:error_function_bounds}) we have that $\mathcal{V}_p$ is lower-bounded by
\begin{equation}
    z^T W_{11} z \leq \mathcal{V}_p
\end{equation}
where $z = [z_1, ~ z_2 ]^T = [\|e_R\|,~\|e_\Omega\|,~ d_\psi({h}_p\|\hat{{h}})]^T \in \mathbb{R}^3$ and $W_{11} \in \mathbb{R}^{3\times3}$ is given by
\begin{equation}
    W_{11} = \begin{bmatrix}
            b_1 k_R & \frac{1}{2}c_2\Lambda_{max}^p & 0 \\
            \frac{1}{2}c_2\Lambda_{max}^p & \frac{1}{2}\Lambda_{min}^p & 0 \\
            0 & 0 & 1
            \end{bmatrix}
\end{equation}
Furthermore, we have
\begin{equation}
    z_1^T W_{13}^p z_1 \leq \mathcal{V}_p \leq z_1^T W_{23}^p z_1
\end{equation}
where $z_1= [\| {e}_R \|,~ \|{e}_\Omega \|]^T$ and $W_{13}^p$, $W_{23}^p \in \mathbb{R}^{2\times2}$ are given by
\begin{equation*}
\small
\begingroup 
\setlength\arraycolsep{0.75pt}
    W_{13}^p = \begin{bmatrix}
        b_1k_R & \frac{1}{2}c_2\Lambda_{max}^p\\
        \frac{1}{2}c_2\Lambda_{max}^p & \frac{1}{2}\Lambda_{min}^p \\
    \end{bmatrix},
    W_{23}^p = \frac{1}{2}\begin{bmatrix}
        b_2k_R & \frac{1}{2}c_2\Lambda_{min}^p \\
        \frac{1}{2}c_2\Lambda_{min}^p & \frac{1}{2}\Lambda_{max}^p \\
    \end{bmatrix}
\endgroup
\end{equation*}
i.e.
\begin{equation}\label{eqn:Vphat_bounds}
    \Lambda_{min}^{W_{13}^p} \| z_1 \|^2 \leq \mathcal{V}_p \leq  \Lambda_{max}^{W_{23}^p} \| z_1 \|^2
\end{equation}
Differentiating $\mathcal{V}_p$ along the solutions of the system and substituting for the control law, $u$, and parameter estimate law, ${\dot{\hat{h}}}$, from (\ref{eqn:control_torqueJhat}) and (\ref{eqn:updaterate_h}) 
\begin{equation}\label{eqn:Vdot}
    \begin{aligned}
    \mathcal{\dot{V}}_p \leq~& -\Big( k_\Omega - \frac{c_2}{\sqrt{2}} \Lambda_{p_{max}} \text{tr}[G]\Big ) \| {e}_\Omega \|^2 - c_2 k_R  \| {e}_R \|^2  \\
    &+ c_2k_\Omega \| {e}_R \| \| {e}_\Omega\|  = -z_1^T W_{31}^p z_1
    \end{aligned}
\end{equation}
where $W_{31}^p \in \mathbb{R}^{2\times2}$ is defined in (\ref{eqn:W31}).
\begin{equation}\label{eqn:W31}
    W_{31} = \begin{bmatrix}
            c_2k_R & -\frac{c_2k_\Omega}{2} \\
            -\frac{c_2k_\Omega}{2} & k_\Omega - \frac{c_2}{\sqrt{2}}\Lambda_{max}^p \text{tr}[G]
            \end{bmatrix}
\end{equation}
\begin{figure}[t]
    \centering
    \includegraphics[width = 0.4\textwidth]{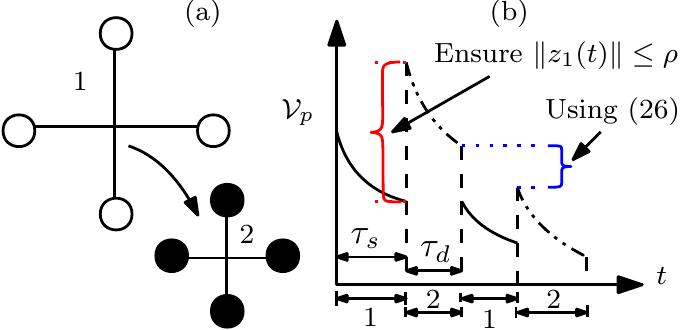}
    \caption{Lyapunov function of the attitude tracking error during configuration switching. $\tau_s$ and $\tau_d$ represent the attitude settling-time and desired dwell-time respectively.}
    \label{fig:lyapunov}
    \vspace{-0.3in}
\end{figure}
This implies that the errors $z_1 = [\|{e}_R \|, ~ \| {e}_\Omega\| ]^T$ asymptotically converge to zero.

\subsubsection*{Remark 6} Although the tracking errors converge to their zero equilibrium, (\ref{eqn:W31}) does not ensure that the parameter errors converge. This is because of the absence of persistence of excitation which would aid in parameter convergence to true values. However, the attitude tracking errors are still guaranteed to be stable and do not depend on the parameter estimation error. 
\rev{\subsubsection*{Remark 7} 
The Assumption 4 requires that the minimum and maximum eigenvalues of the true inertia matrix be known. These values are only used to find the constant $c$ in (\ref{eqn:updaterate_h}) and therefore can be relaxed such that values from approximate CAD models should be enough \cite{PM+20,PM+21}}. 
\subsubsection{Stability of the switched system}
We will again use multiple Lyapunov functions to establish the stability of the attitude tracking dynamics with the proposed adaptive controller. Consider the following \textit{Proposition 8}:
\subsubsection*{Proposition 8}\label{sec:prop6} Consider the system (\ref{eqn:ps}) and that Assumption 4 holds. With the control generated according to (\ref{eqn:control_torqueJhat})-(\ref{eqn:updaterate_h}), if the initial guess of inertia parameters, $\hat{{h}}_p$ for each subsystem is adaptively updated and the switching is performed at time $t_j >> t_i$ such that (\ref{eqn:switch_condition}) holds, then the attitude tracking errors, $e_{R},e_\Omega$, of the switched system converge to zero asymptotically.
 \begin{equation}\label{eqn:switch_condition}
     \| z_1(t_j) \|^2 \leq \Bigg ( \frac{\Lambda_{min}^{W_{13}^p}}{\Lambda_{max}^{W_{23}^p}} \Bigg ) \| z_1(t_i) \|^2
 \end{equation}

\subsubsection*{Proof}
To analyze this case, consider a switched system generated by two dynamical systems such that $\mathcal{P} = [1, ~2]$. Let $t_i < t_j$ be two switching times when $\sigma = 1$.
Then, using \textit{Proposition 8}, the fourth term in (\ref{eqn:lyapunov_p}), $d_\psi (\boldsymbol{h}_p \| \boldsymbol{\hat{h}})$, is adaptively updated from the previous value, hence is constant at the two time instants $t_i$ and $t_j$. Next, (\ref{eqn:Vphat_bounds}) provides the bounds on the first three terms of the Lyapunov candidate at the two time intervals. Hence if the switching time instant is chosen such that (\ref{eqn:switch_condition}) holds, the switched system is asymptotically stable using \textit{Lemma 2}.

\subsubsection*{Remark 9} Note that \textit{Proposition 8} enforces the minimum dwell-time ($\tau_d$) requirement for the switched system stability. As mentioned in Remark 3, the planner is made aware of the dwell-time such that the reference trajectory is generated to accommodate the dwell-time requirements as described in the following Section. \ref{sec:planning}.

\subsubsection*{Remark 10} Since it is well-known that the adaptive controllers can be unstable even for slight disturbance, we modify the control law proposed in (\ref{eqn:control_torqueJhat})-(\ref{eqn:updaterate_h}) to include a robust term in the following Section \ref{sec:caseC}.
\vspace{-0.05in}
\rev{\subsection{Case with model uncertainties in $\textit{{H}}_\textit{p}$ and  external disturbances, $\Delta \neq [0 ~0 ~0]^T$}\label{sec:caseC}
Finally we discuss the case when we have modelling uncertainties coupled with external disturbances to improve the robustness of the proposed adaptive controller in the presence of disturbances.}
\begin{figure*}[t]
    \centering
    \subfloat[Tracking of angular velocity ($\Omega$:red solid, $\Omega_d$:blue dashed)]{\includegraphics[width = 0.48\textwidth]{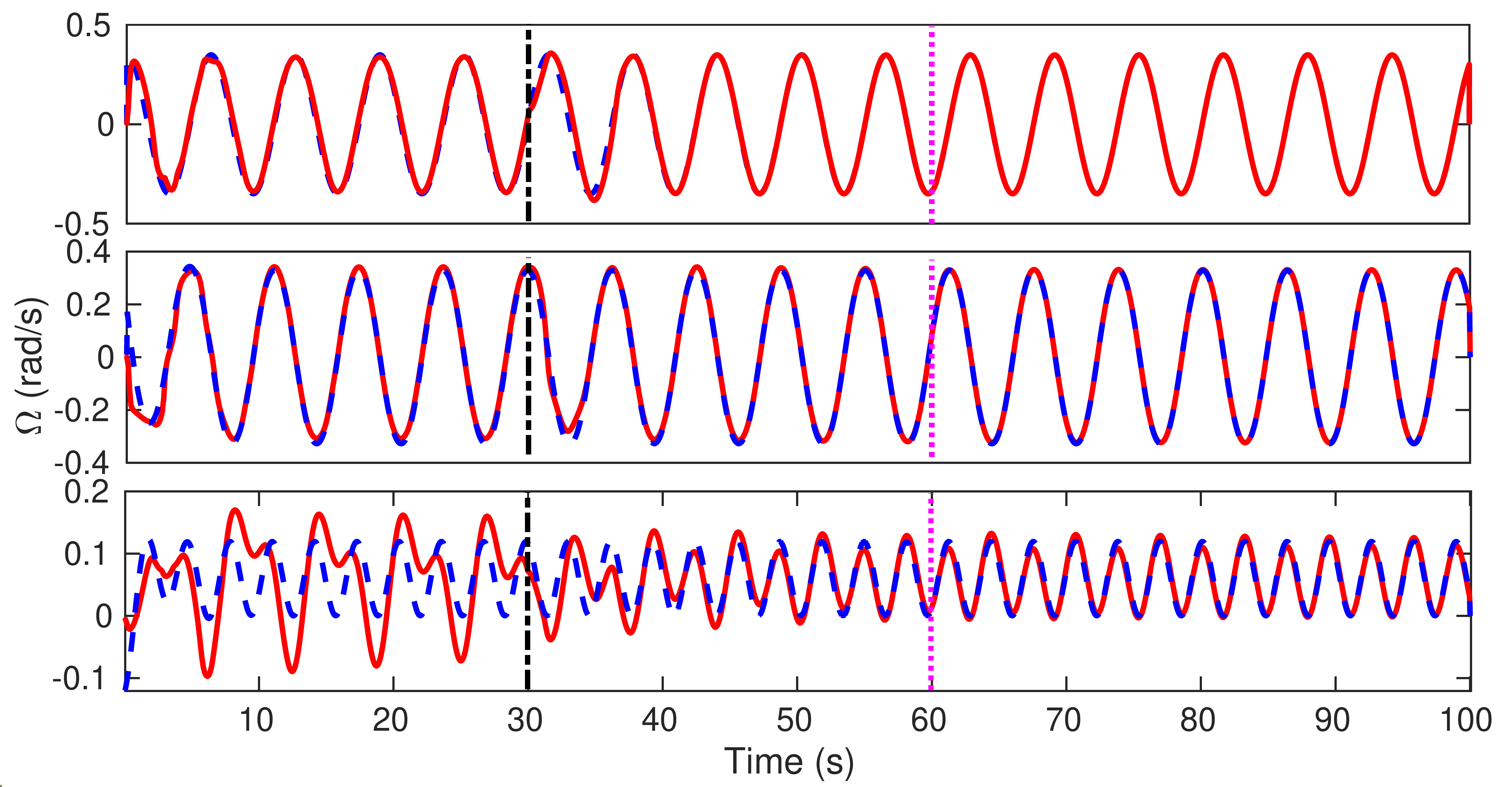}}
    \subfloat[Errors in $\boldsymbol{e}_R$ ($e_R$:red solid, Reference:blue dashed)]{\includegraphics[width = 0.48\textwidth]{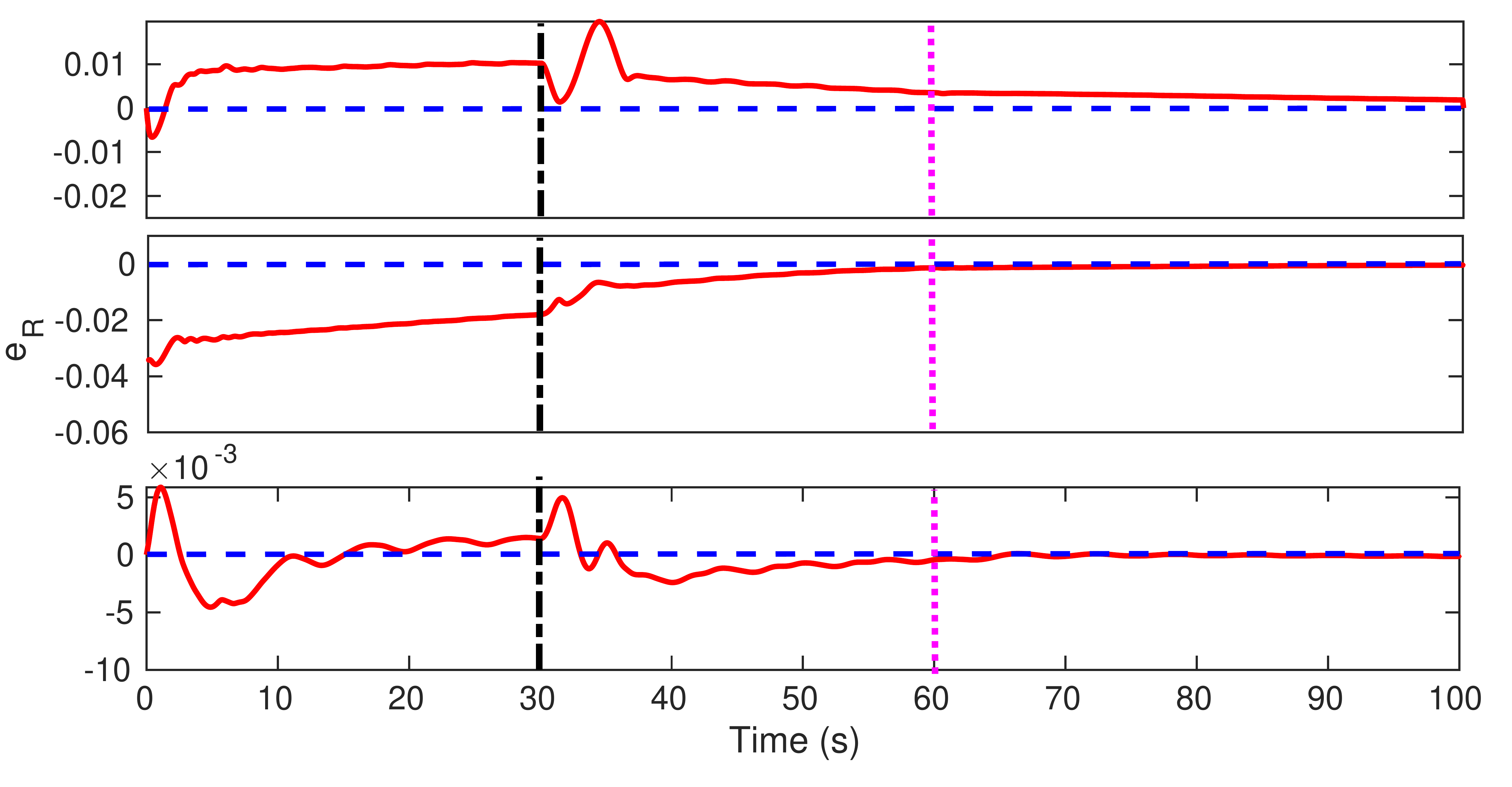}}\\
    \vspace{-0.1in}
    \subfloat[Inertia estimates (kg-m$^2$) (true: dotted, estimated: solid)]{\includegraphics[width = 0.48\textwidth]{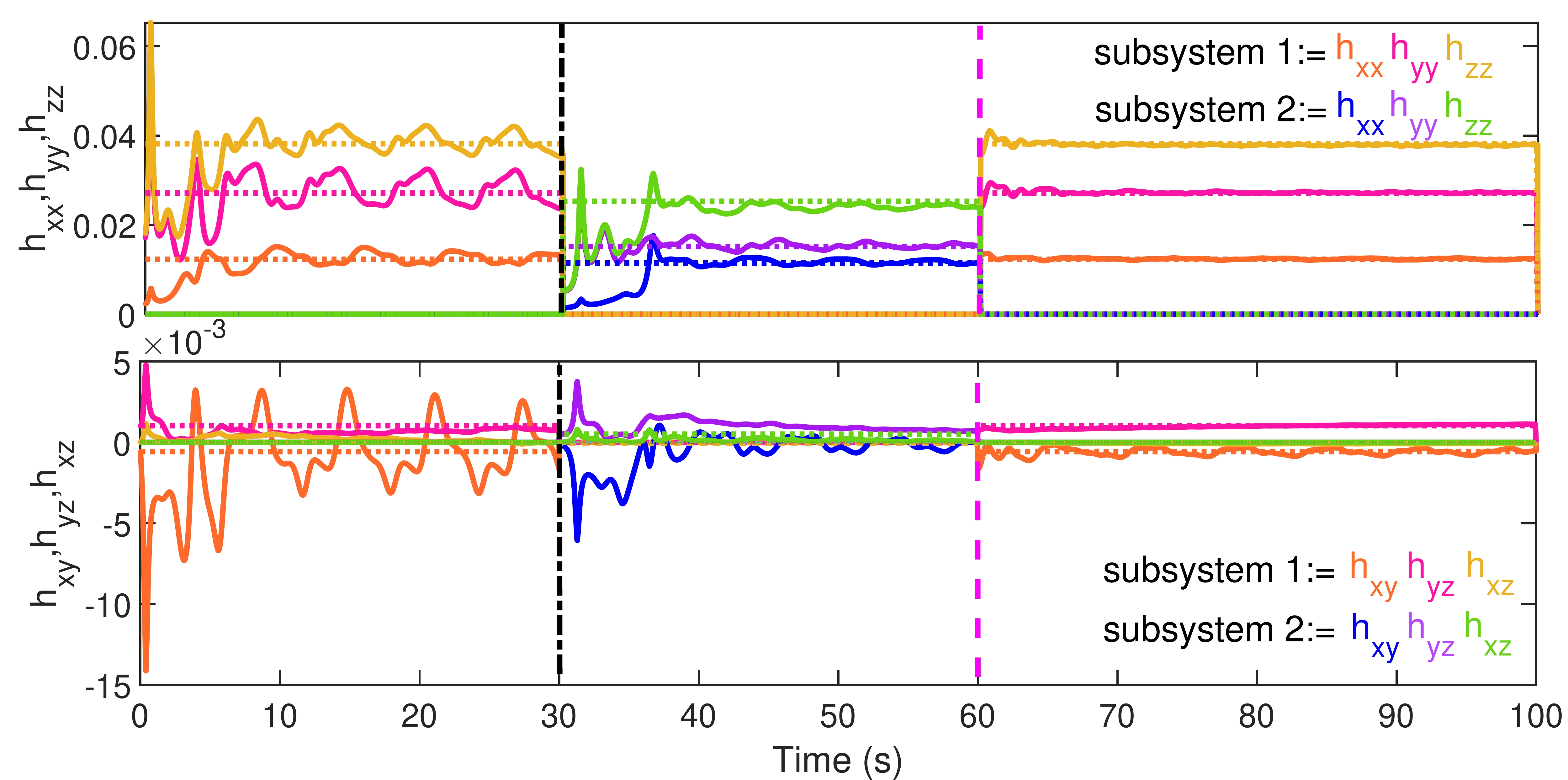}}
    \subfloat[Control effort ]{\includegraphics[width = 0.48\textwidth]{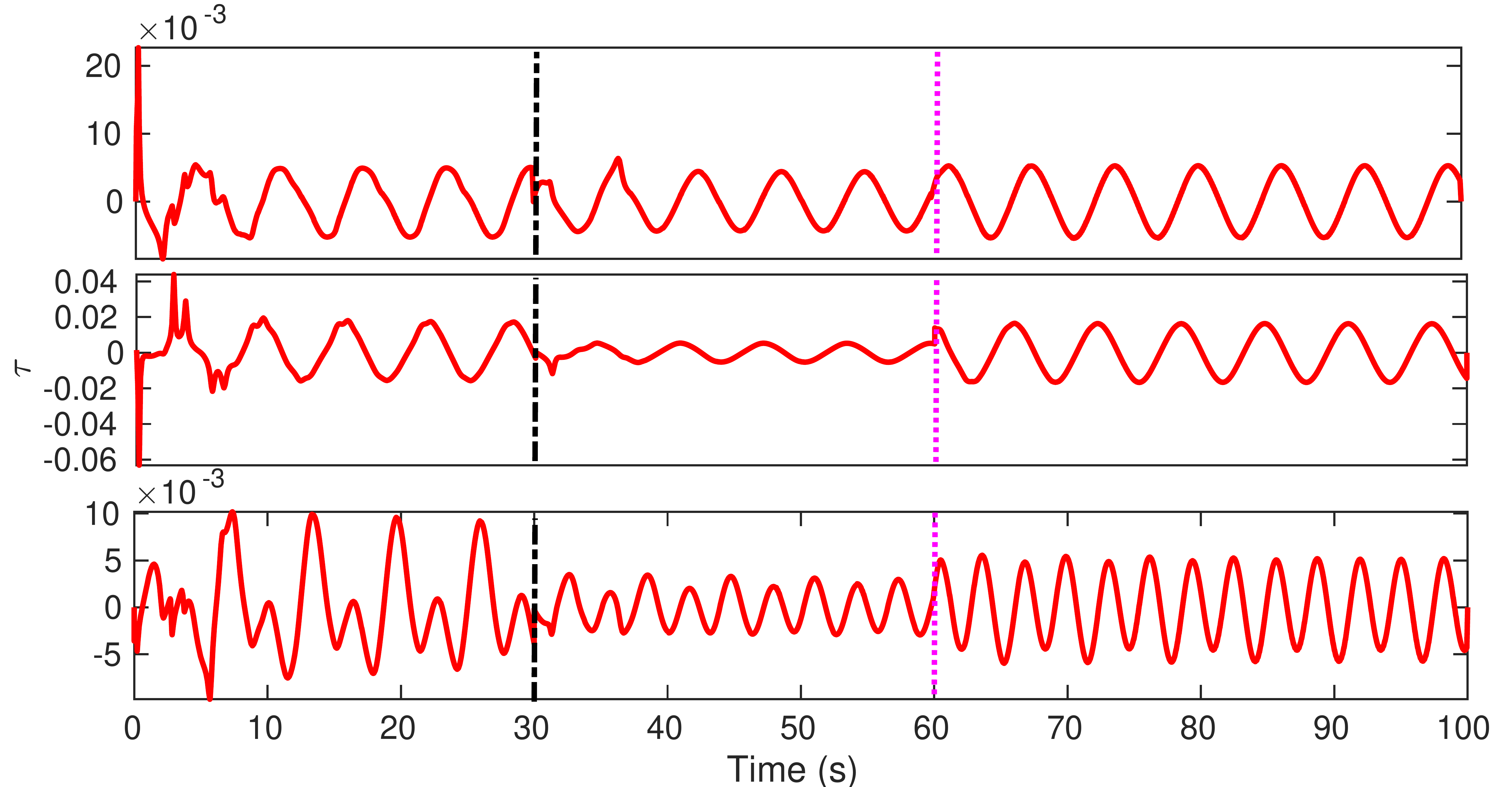}}
    \caption{Performance of the proposed attitude controller when the vehicle switches between the two configurations shown in Fig. \ref{fig:lyapunov}(a) at $t = 30s$ from 1 to 2 (black dash-dotted vertical line) and at $t= 60s$ from 2 to 1 (pink dashed vertical line) by following (\ref{eqn:switch_condition}). With the proposed adaptive controller, the attitude errors converge to the reference (horizontal dotted lines).}
    \label{fig:adaptive_controller}
    \vspace{-0.3in}
\end{figure*}
\rev{\subsubsection*{Assumption 11} The disturbances in attitude dynamics have known bounds, i.e, $\| \Delta \| \leq \delta_R$ for a positive constant.} 
\rev{\subsubsection*{Proposition 12} Suppose Assumptions 4 and 10 hold. Then, if the control torques are generated according to
\begin{align}
        u &= -k_R {e}_R - k_\Omega {e}_\Omega - {Y}{\hat{h}} + \mu, \\\label{eqn:control_torqueSMC}
    {\dot{\hat{h}}} &=
        - (\nabla^2\psi({\hat{h}}))^{-1} {Y}^T {e}_A,  \\ 
        \mu &= - \Big(\delta_R - \frac{\eta}{ \norm{e_A}}\Big) \frac{e_A}{\norm{e_A}},\\
        e_A &= e_\Omega + c_2 e_R
\end{align}
where $\eta$ is a small positive constant which is adaptively chosen such that $\eta < z_1^T W_{31}^p z_1 $, the attitude tracking errors asymptotically converge to their zero equilibrium.}
\rev{\subsubsection*{Proof} The proof is similar as presented in Section \ref{sec:caseB} and is given in the \href{https://arxiv.org/abs/2209.08676}{Appendix} Section D}.
\rev{\subsubsection*{Remark 13} The Assumption 11 assumes that the disturbances in the attitude dynamics are bounded \cite{LS21,tLee}. Since this value is used to generate the robust control term $\mu$, a rough approximate can be used based on the aerodynamic conditions of the flight space.}
\section{Control-Aware Minimum Jerk Trajectory}\label{sec:planning}
The \textit{Proposition 8} in \ref{sec:prop6} with (\ref{eqn:switch_condition}) implies that for the switched system to have asymptotic tracking
stability, the minimum dwell-time before switching should be calculated as a percentage of the initial tracking error. This is shown by the blue line in Fig. \ref{fig:lyapunov}. However, this doesn't still quantify the bounds of $\| \mathcal{V}_i(t) - \mathcal{V}_j(t) \|~ \forall i \neq j; i,j \in \mathcal{P}$ as shown by the red line. 
\subsubsection*{Assumption 14} The upper bound on the estimation error for the $p^{th}$ subsystem, $z_{2}^p(t)$ is known.
\subsubsection*{Assumption 15} The settling time corresponding to the maximum attitude error for the $p^{th}$ subsystem is known.
\subsubsection*{Proposition 16} Suppose that Assumptions 4,  14 and 15 hold, if switching is performed at $t = \tau_s$ when $\| z_1^p(\tau_s) \| \leq \rho $ where $\tau_s$ denotes the settling-time for the attitude errors, $e_R$ and $e_\Omega$, and $\rho > 0$ denotes the region within which the errors remain, we have the minimum value of the difference in the two Lyapunov functions at the same time instant (the jump in the Lyapunov value, shown by the red line in Fig. \ref{fig:lyapunov}b), as 
\begin{align}\label{eqn:VtBounds}
    \| \mathcal{V}_i(\tau_s) -  \mathcal{V}_j(\tau_s)  \|
\leq  ~& (\Lambda_{max}^{W_{21}^i} + \Lambda_{max}^{W_{21}^j})\rho \\
&+
\Lambda_{max}^{W_{21}^i}\|z_{2}^i(\tau_s) \| + \Lambda_{max}^{W_{21}^j}\| z_{2}^j(\tau_s) \| \nonumber
\end{align}
\subsubsection*{Proof} \textit{Proposition 16} directly follows from the minimum value of (\ref{eqn:boundVp}) by employing (\ref{eqn:lyapunov_p}):
\begin{equation}\label{eqn:boundVp}
\small
    \mathcal{V}_p \leq  \Lambda_{max}^{W_{21}^p} \| z \|^2, \text{ with }
        W_{21} = \begin{bmatrix}
            b_2 k_R & \frac{1}{2}c_2\Lambda_{min}^p & 0 \\
            \frac{1}{2}c_2\Lambda_{min}^p & \frac{1}{2}\Lambda_{max}^p & 0 \\
            0 & 0 & 1
            \end{bmatrix}
\end{equation}
\vspace{-0.1in}
\rev{\subsubsection*{Remark 17} The Assumption 14 implies that the maximum estimation error should be bounded. This can be achieved in various ways for example by using the Projection operator, \cite{tsakalis1998,IS12}, and by assuming a maximum estimation error.}
\vspace{-0.02in}
\rev{\subsubsection*{Remark 18} The Assumption 15 requires that the settling time for the quadrotor for a $p^{th}$ configuration be known and this information can be approximated estimated as a rough upper bound from real experimental data.}

Since the position controller is a proportional-derivative control on position, waypoint planning to fly through passages is not ideal which would result in high initial attitude errors such that $z_1^i(t) \nleq \rho$ if the vehicle switches before the attitude errors' settling-time. Alternatively, the minimum-jerk trajectory (MJT) planner can be successfully employed here to ensure $z_1^i(\tau) \leq \rho$ by imposing the desired velocity boundary conditions at the entrance of the passageway, where configuration switching is mandated by the geometric constraint. The time taken to reach this velocity should be set to at least $\tau_s$. By ensuring that the vehicle has attained this velocity, the attitude errors will be lower at the entrance of the passageway in the absence of external disturbances. Hence this will lead to lower bounds on the tracking errors as given by (\ref{eqn:VtBounds}).

The MJT planner is generated according to
\begin{equation}
    r^*(t) = \underset{r(t)}{argmin} \int_0^T \dddot{r}^2 ~dt
\end{equation}
with the following boundary conditions:
\begin{equation}
\centering
    \begin{aligned}
        r(0) = [0,0,0]^T,& ~\dot{r}(0) = [0,0,0]^T, ~\ddot{r}(0) = [0,0,0]^T \\
        r(\tau) = r_{des},&~ \dot{r}(\tau) = \dot{r}_{des}, ~\ddot{r}(\tau) = [0,0,0]^T
    \end{aligned}
\end{equation}
where $r_{des}$ and $\dot{r}_{des}$ denote the coordinates of the entrance of the passageway and the desired velocity to fly through the passageway respectively and $\tau \geq \text{max }\{\tau_s,\tau_d\}$ where $\tau_d$ is defined as $t_j - t_i$ from (\ref{eqn:switch_condition}) for the $p^{th}$ system.

\section{Results and Discussion}\label{sec:results}
This section describes the various case scenarios simulated to validate the proposed controller for the switched system. The position controller from Fig. \ref{fig:firstpic} is implemented from \cite{L+10} to generate the necessary desired orientation and thrust. 
Please refer to Appendix Section I1 for further details about the simulation parameters.
Results for the case scenario (2) are shown in Fig. \ref{fig:adaptive_controller}(a)-(d). We show how the switching is performed after the errors have decreased and the tracking errors converge to their zero equilibrium to validate \textit{Proposition 8} in the presence of modeling uncertainties in inertia. The parameter estimates also converge (however, this is not guaranteed due to the absence of persistence of excitation). 
Since the developed controller is a PD controller, it is inherently robust to small uncertainties and hence almost perfect tracking of roll and pitch rate in the body frame is observed even when the inertia estimates oscillate. However, for the subsystem 1, the uncertainty in the $z$ direction is significantly higher, implying the yaw torque was not enough. Yaw rate tracking is eventually achieved by utilizing the estimation of the inertia parameters. 
Additional validation results for case (3) and the comparison plots with a conventional robust controller are presented in the Sections I2-I3 of \cite{appendix}.

\begin{figure}[t]
    \centering
    \includegraphics[width = 0.45\textwidth]{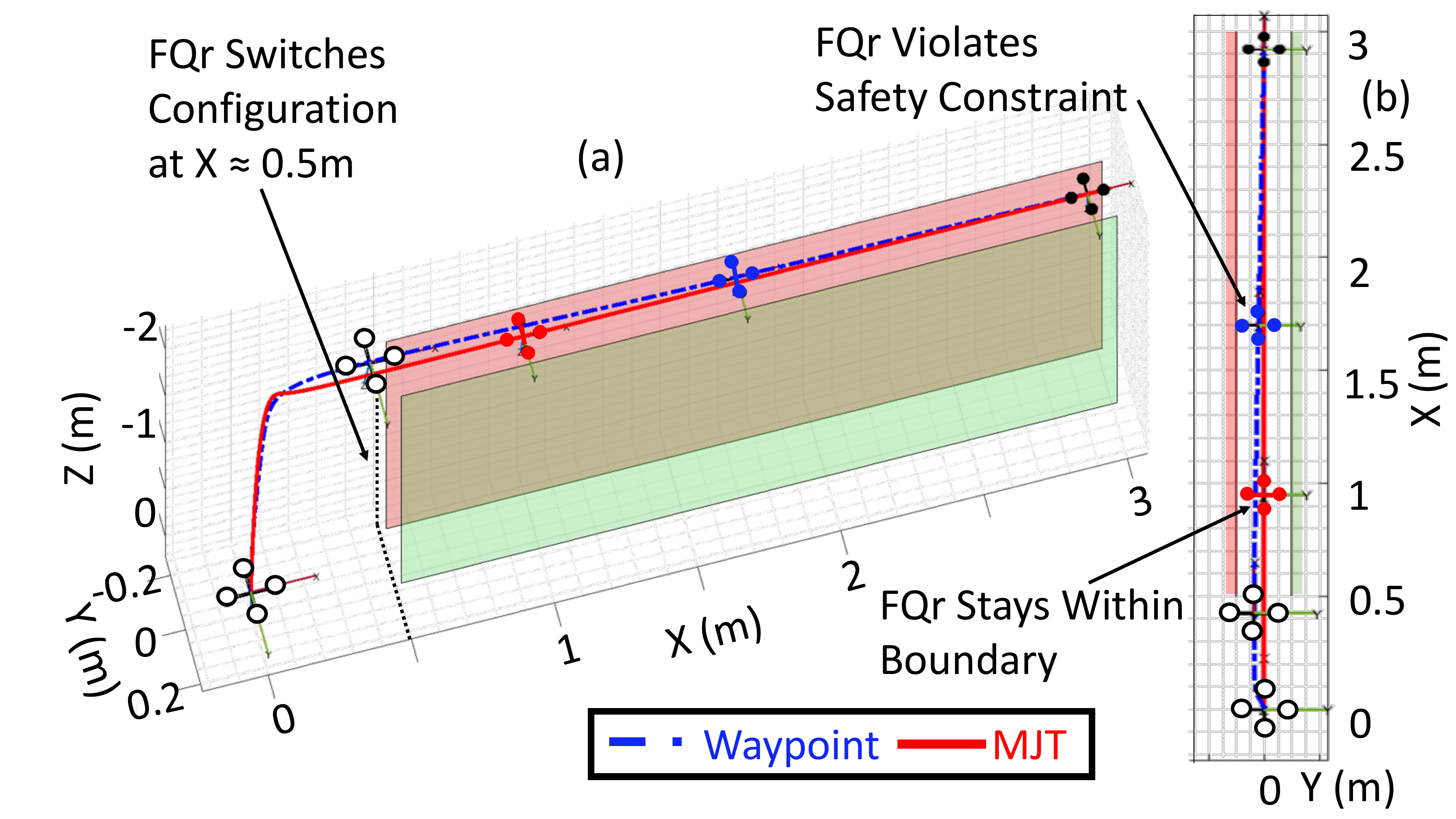}
    \vspace{-0.1in}
    \caption{The tracking results for minimum-jerk trajectory and waypoint based methods. The MJT based trajectory leads to low deviations from the trajectory, while the waypoint based one leads to safety constraint violations.}
    \label{fig:tracking}
    \vspace{-0.3in}
\end{figure}

Next, we integrate the proposed attitude controller with a minimum-jerk trajectory planner and compare the performance against a waypoint-based planner to validate \textit{Proposition 12}. The MJT-based planning framework demonstrates how the vehicle transitions from the initial configuration to the new configuration at $[0.5~0 ~-2]^T$m at $t = 9.02s$ without giving rise to additional tracking errors as shown in Fig. \ref{fig:tracking}(a)-(b), shown in red solid lines. The waypoint based planner, however, arrives at the same position at $t=5.24s$ which is less than the maximum attitude settling time ($\tau_s = 8.87s)$ and therefore has high attitude errors during the transitioning. This leads to higher switch-based disturbances, violating the safety constraints as shown in \ref{fig:tracking}(b).

\vspace{-0.04in}
\section{Conclusion}\label{sec:conclusion}
In this article, we presented an approach for analyzing the attitude tracking stability of foldable quadrotors (FQrs) by modeling them as switched systems. We employed this analysis to design an adaptive control law and derived the necessary dwell-time requirements for guaranteeing the asymptotic stability of the attitude tracking errors in the presence of bounded disturbances. 
Another highlight of the work was to exploit the attitude settling-time information and design the boundary conditions for a control-aware trajectory planner to achieve stable flights during switching. \rev{Future work includes extension of the adaptive control law to account for other matched and mismatched input uncertainties.}

\bibliography{bibliography.bib}
\bibliographystyle{ieeetr}

\appendix
\section{Notations and Definitions}
\subsection{Switching signal}
A piece-wise constant function $\sigma : [0,\infty) \rightarrow \mathcal{P}.$ with a finite number of discontinuities and takes a constant value on every interval between two consecutive switching time instants. The role of $\sigma$ is to specify, at each time instant $t$, the index $\sigma(t) \in \mathcal{P}$ of the active subsystem model from the family of switched systems \cite{liberzon}.
\subsection{Hat map}
The \textit{hat map} $\hat{\cdot}: \mathbb{R}^3 \xrightarrow{} \mathsf{SO(3)}$ is a symmetric matrix operator defined by the condition that $\hat{x}y = x \times y ~\forall~ x,y \in \mathbb{R}^3$. An example is provided here:\\
The angular velocity vector ${\Omega }=[\omega _{x},\omega _{y},\omega _{z}]^T$ may be equivalently expressed as an angular velocity tensor, the matrix (or linear mapping) defined by:
\begin{equation}
    \begin{aligned}
        \hat{\Omega} = \begin{pmatrix}
        0&-\omega _{z}&\omega _{y}\\\omega _{z}&0&-\omega _{x}\\-\omega _{y}&\omega _{x}&0\\
        \end{pmatrix}
    \end{aligned}
\end{equation}
\subsection{Vee map}
The \textit{vee map} $\vee: \mathsf{SO(3)} \xrightarrow{} \mathbb{R}^3$ represents the inverse of the hat map and $[.]_\times$ is the skew symmetric cross product matrix. For example, $\hat{\Omega}^\vee = \Omega$. 
\subsection{Attitude dynamics and $Y_1$}
The attitude dynamics for an individual subsystem can be rewritten in the form of
\begin{equation}
    {H_p}{\dot{\Omega}} - {Y}_1 {h}_p  = {u} + \Delta
\end{equation}
where ${Y}_1$ is given by 
\begin{equation}\label{eqn:y1}
\begingroup 
\setlength\arraycolsep{1.5pt}
    {Y}_1 = \begin{bmatrix}
        0 & \omega_2\omega_3 & -\omega_2\omega_3 & \omega_1 \omega_3 & -\omega_1 \omega_2 & \omega_3^2 - \omega_2^2 \\
        - \omega_1 \omega_3 & 0 & \omega_1 \omega_3 & -\omega_2 \omega_3 & \omega_1^2 - \omega_3^2 & \omega_2 \omega_1 \\
         \omega_1 \omega_2 & -\omega_2 \omega_1 & 0 & \omega_2^2 -\omega_1^2 & \omega_3 \omega_2 & -\omega_3 \omega_2 
    \end{bmatrix}
\endgroup
\end{equation}
and 
\[{h}_p =  \begin{bmatrix}h_{p_{xx}} \\h_{p_{yy}} \\h_{p_{zz}} \\h_{p_{xy}} \\ h_{p_{xz}} \\ h_{p_{zz}} \end{bmatrix}\] 
is the vector encompassing the unique elements of the moment of inertia tensor.

\subsection{Definition of $Y_2$}
$Y_2$ is defined as 
\begin{equation}\label{eqn:y2}
    {Y}_2 = \begin{bmatrix}
    \alpha_{d1} & 0& 0& \alpha_{d2}& \alpha_{d3}& 0 \\
    0&  \alpha_{d2}& 0& \alpha_{d1} &0 &  \alpha_{d3} \\
    0& 0&  \alpha_{d3}& 0&  \alpha_{d1}&  \alpha_{d2} \\
    \end{bmatrix}
\end{equation} 
\subsection{Case (1)}
In this section, we present the detailed proof for the exponential convergence of tracking errors for individual subsystems. The control moment in this case is generated according to (\ref{eqn:control_tau_1}) as proposed in \cite{L+10}. 
\begin{align}
        u &= -k_R {e}_R - k_\Omega {e}_\Omega - {Y}_1{h}_p  \label{eqn:control_tau_1_app} 
\end{align}
where ${Y} = {Y}_1 - {Y}_2$.
\subsection*{Proof}
Consider the individual subsystem's Lyapunov candidate $\forall p = 1,2..m \in \mathcal{P}$ as
\begin{equation}
\begin{aligned}
          \mathcal{V}_p =~& \frac{1}{2}e_\Omega^T {H_p}e_\Omega + k_R\Phi(R,R_d) + c_1 {e}_R \cdot {H_p}  {e}_\Omega 
\end{aligned}
\end{equation}
In the sub-level set $\mathcal{L}$, we have
\begin{equation}
    z^T W_1^p z \leq \mathcal{V}_p \leq z^T W_2^p z
\end{equation}
where $z= [\| {e}_R \|~ \|{e}_\Omega \|]^T$ and $W_1^p$, $W_2^p$ are given by
\begin{equation*}
	\setlength\arraycolsep{0.5pt}
    W_1^p = \begin{bmatrix}
        b_1k_R &\frac{1}{2}c_1\Lambda_{p_{max}} \\
        \frac{1}{2}c_1\Lambda_{p_{max}} & \frac{1}{2}\Lambda_{p_{min}} \\
    \end{bmatrix},
    W_2^p = \begin{bmatrix}
        b_2k_R &\frac{1}{2}c_1\Lambda_{p_{max}} \\
        \frac{1}{2}c_1\Lambda_{p_{max}} & \frac{1}{2}\Lambda_{p_{min}} \\
    \end{bmatrix}
\end{equation*}
i.e.,
\begin{equation}\label{eqn:Vbounds}
    \Lambda_{min}^{W_1^p} \| z \|^2 \leq \mathcal{V}_p \leq  \Lambda_{max}^{W_2^p} \| z \|^2
\end{equation}
Differentiating $\mathcal{V}$ along the solutions of the system 
\begin{align*}
    \mathcal{\dot{V}}_p =~& {e}_\Omega^T {H_p} {\dot{e}}_\Omega + k_R {e}_R \cdot {e}_\Omega + c_1 {\dot{e}}_R \cdot {H_p} {e}_\Omega + c_1 {e}_R \cdot {H_p} {\dot{e}}_\Omega  \\
    \end{align*}
    Now, substituting (\ref{eqn:y1}),(\ref{eqn:y2}) and  (\ref{eqn:control_tau_1_app}), we obtain
    \begin{align*}
    =~&  {e}_\Omega ^T (-k_R {e}_R - k_\Omega {e}_\Omega - {Y}{h}_p + {Y}_1 {h}_p ) \\
    &- {e}_\Omega ^T {Y}_2 {h}_p + k_R {e}_R \cdot {e}_\Omega +  c_1 C(R_d^T,R){e}_\Omega \cdot {H_p} {e}_\Omega \\
    &+ c_1 {e}_R^T (-k_R {e}_R - k_\Omega {e}_\Omega + {Y}{\hat{h}}_p + {Y}_1 {h}_p)-c_1 {e}_R^T {Y}_2 {h}_p \\
     =~& -k_\Omega {e}_\Omega ^T {e}_\Omega +  c_1 C(R,R_d){e}_\Omega \cdot {H_p} {e}_\Omega - c_1 k_R {e}_R^T {e}_R \\
     &- c_1k_\Omega {e}_R \cdot {e}_\Omega.
\end{align*}
Since $C(R,Rd)e_\Omega \leq \frac{1}{\sqrt{2}}\text{tr}[G] \|e_\Omega \|$, 
\begin{equation}
    \begin{aligned}
    \mathcal{\dot{V}}_p 
    \leq~& -\Big( k_\Omega - \frac{c_1}{\sqrt{2}} \Lambda_{p_{max}} \text{tr}[G]\Big ) \| {e}_\Omega \|^2 - c_1 k_R  \| {e}_R \|^2  \\
    &+ c_1k_\Omega \| {e}_R \| \| {e}_\Omega\| = -z^T W_3 z,
    \end{aligned}
\end{equation}
where $W_3^p$ is given by (\ref{eqn:W3}) 
\begin{equation}\label{eqn:W3}
    W_3^p = \begin{bmatrix}
        c_1k_R & -\frac{c_1}{2} \\
        -\frac{c_1}{2} & k_\Omega - \frac{c_1}{\sqrt{2}}\Lambda_{p_{max}}\text{tr}[G]. 
    \end{bmatrix}
\end{equation}
Therefore, 
\begin{equation}\label{eqn:VdotBounds}
    \dot{\mathcal{V}}_p \leq -\Lambda_{min}^{W_3^p} \| z \|^2
\end{equation}
Let $\beta_p = \frac{\Lambda_{min}^{W_3^p}}{2\Lambda_{max}^{W_2^p}}$, then from (\ref{eqn:Vbounds} and (\ref{eqn:VdotBounds}), we have
\begin{equation}
    \dot{\mathcal{V}}_p \leq -2\beta_p \mathcal{V}_p
\end{equation}
Hence the tracking errors are exponentially stable for the individual subsystems. This implies that if $\sigma(t) = p$ for $t \in [t_0,t_0 + \tau_d)$, we have
\begin{equation}
    \mathcal{V}_p (z(t_0 + \tau_d)) \leq e^{-2\beta_p\tau_d}\mathcal{V}_p(z(t_0))
\end{equation}

\subsection{Case (2)}
In this section, we present the proof for asymptotic stability of tracking errors for individual subsystems when the inertia parameters are not known and there are no external disturbances. The control torques are generated according to 
\begin{equation}\label{eqn:control_torqueJhat_app}
    \begin{aligned}
        u &= -k_R {e}_R - k_\Omega {e}_\Omega - {Y}{\hat{h}}_p  \\
    {\dot{\hat{h}}}_p &=
        - (\nabla^2\psi({\hat{h}}_p))^{-1} {Y}^T {e}_A, \\
        e_A &= e_\Omega + c_2 e_R
\end{aligned}
\end{equation}
\subsection*{Proof}
\vspace{-0.2in}
\begin{equation}\label{eqn:lyapunov_p_app}
\begin{aligned}
          \mathcal{V}_p =~& \frac{1}{2}e_\Omega^T {H_p}e_\Omega + k_R\Phi(R,R_d) + c_2 {e}_R \cdot {H_p}  {e}_\Omega + d_\psi ({h}_p \| {\hat{h}})
\end{aligned}
\end{equation}
where $ d_\psi ({h}_p \| {\hat{{h}}_p})$ is the Bregman divergence operator \cite{L+18}:
\begin{equation*}
\begin{aligned}
    d_\psi ({h}_p \| {\hat{{h}}_p}) = \psi({h}_p) - \psi(\hat{{h}}_p) - ({h}_p - \hat{{h}}_p)^T\nabla \psi(\hat{{h}}_p)
\end{aligned}
\end{equation*}
and the time-derivative of $d_\psi ({h}_p \| {\hat{h}}_p)$ is
\begin{equation}
\begin{aligned}
          \dot{d_\psi }(\cdot) & =
        (\hat{{h}}_p - {h}_p)^T\nabla^2 \psi(\hat{{h}}_p)\dot{\hat{{h}}}_p
\end{aligned}
\end{equation}
As shown in \cite{L+18}, $d_\psi ({h}_p \| {\hat{h}}_p)$ can be taken as an approximation for the geodesic estimation error with the properties required of a desired Lyapunov candidate. Also, from \[ b_1 \| e_R(R,Rd) \|^2 \leq \Phi(R,Rd) \leq b_2 \| e_R(R,Rd) \|^2 \] we have that $\mathcal{V}_p$ is lower-bounded by
\begin{equation}
    z^T W_{11} z \leq \mathcal{V}_p
\end{equation}
where $z = [z_1, ~ z_2 ]^T = [\|e_R\|,~\|e_\Omega\|,~ d_\psi({h}_p\|\hat{{h}})]^T \in \mathbb{R}^3$ and $W_{11} \in \mathbb{R}^{3\times3}$ is given by
\begin{equation}
    W_{11} = \begin{bmatrix}
            b_1 k_R & \frac{1}{2}c_2\Lambda_{max}^p & 0 \\
            \frac{1}{2}c_2\Lambda_{max}^p & \frac{1}{2}\Lambda_{min}^p & 0 \\
            0 & 0 & 1
            \end{bmatrix}
\end{equation}
Furthermore, we have
\begin{equation}
    z_1^T W_{13}^p z_1 \leq \mathcal{V}_p \leq z_1^T W_{23}^p z_1
\end{equation}
where $z_1= [\| {e}_R \|,~ \|{e}_\Omega \|]^T$ and $W_{13}^p$, $W_{23}^p \in \mathbb{R}^{2\times2}$ are given by
\begin{equation*}
	\setlength\arraycolsep{0.5pt}
    W_{13}^p = \begin{bmatrix}
        b_1k_R & \frac{1}{2}c_2\Lambda_{max}^p\\
        \frac{1}{2}c_2\Lambda_{max}^p & \frac{1}{2}\Lambda_{min}^p \\
    \end{bmatrix},
    W_{23}^p = \frac{1}{2}\begin{bmatrix}
        b_2k_R & \frac{1}{2}c_2\Lambda_{min}^p \\
        \frac{1}{2}c_2\Lambda_{min}^p & \frac{1}{2}\Lambda_{max}^p \\
    \end{bmatrix}
\end{equation*}
i.e.
\begin{equation}\label{eqn:Vphat_bounds_app}
    \Lambda_{min}^{W_{13}^p} \| z_1 \|^2 \leq \mathcal{V}_p \leq  \Lambda_{max}^{W_{23}^p} \| z_1 \|^2
\end{equation}
Differentiating $\mathcal{V}_p$ along the solutions of the system and employing (\ref{eqn:control_torqueJhat_app}), we obtain
\begin{align*}
    \mathcal{\dot{V}}_p =& {e}_\Omega^T {H_p} {\dot{e}}_\Omega + k_R {e}_R \cdot {e}_\Omega + c_2 {\dot{e}}_R \cdot {H_p} {e}_\Omega \\
    &+ c_2 {e}_R \cdot {H_p} {\dot{e}}_\Omega + \dot{d_\psi }(\cdot) \\
     =& -k_\Omega {e}_\Omega ^T {e}_\Omega + {e}_\Omega ^T {Y}({\hat{h}}_p - {h}_p) +
      c_2 C(R_d^T,R){e}_\Omega \cdot {H_p} {e}_\Omega \\
      & - c_2 k_R {e}_R^T {e}_R + c_2 {e}_R \cdot {Y}({\hat{h}}_p - {h}_p)  -c_2k_\Omega {e}_R \cdot {e}_\Omega + \dot{d_\psi }(\cdot) \\
     =& -k_\Omega {e}_\Omega ^T {e}_\Omega +  c_2 C(R_d^T,R){e}_\Omega \cdot {H_p} {e}_\Omega - c_2 k_R {e}_R^T {e}_R \\
     &+ {e}_A \cdot {Y}({\hat{h}}_p - {h}_p) - c_2 k_\Omega {e}_R \cdot {e}_\Omega + \dot{d_\psi }(\cdot)
\end{align*}

Substituting for the control law, $u$, and parameter estimate law, $\dot{\hat{h}}$ 
\begin{equation}\label{eqn:Vdot}
    \begin{aligned}
    \mathcal{\dot{V}}_p =~& -k_\Omega {e}_\Omega ^T {e}_\Omega  - c_2 k_R {e}_R^T {e}_R + c_2 C(R_d^T,R){e}_\Omega \cdot {H_p} {e}_\Omega
    \\
     &- c_2 k_\Omega {e}_R \cdot {e}_\Omega \\
    \leq~& -\Big( k_\Omega - \frac{c_2}{\sqrt{2}} \Lambda_{p_{max}} \text{tr}[G]\Big ) \| {e}_\Omega \|^2 - c_2 k_R  \| {e}_R \|^2  \\
    &+ c_2 k_\Omega \| {e}_R \| \| {e}_\Omega\| = -z_1^T W_{31}^p z_1
    \end{aligned}
\end{equation}
where $W_{31}^p \in \mathbb{R}^{2\times2}$ is defined in (\ref{eqn:W31_app}).
\begin{equation}\label{eqn:W31_app}
    W_{31} = \begin{bmatrix}
            c_2 k_R & -\frac{c_2 k_\Omega}{2} \\
            -\frac{c_2 k_\Omega}{2} & k_\Omega - \frac{c_2}{\sqrt{2}}\Lambda_{max}^p \text{tr}[G]
            \end{bmatrix}
\end{equation}

This implies that the errors $z_1 = [\|{e}_R \|, ~ \| {e}_\Omega\| ]^T$ asymptotically converge to zero.

\subsection{Case (3)}
In this section, we present the proof for the asymptotic stability of the attitude tracking errors of the $p^{th}$ subsystem in the presence of modeling uncertainty and external disturbances. The control torques are generated according to
\begin{equation}\label{eqn:control_torqueSMC_app}
    \begin{aligned}
        u &= -k_R {e}_R - k_\Omega {e}_\Omega - {Y}{\hat{h}} + \mu , \\
    {\dot{\hat{h}}} &=
        - (\nabla^2\psi({\hat{h}}))^{-1} {Y}^T {e}_A,  \\ 
        \mu &= - \Big(\delta_R + \frac{\eta}{ \norm{e_A}}\Big) \frac{e_A}{\norm{e_A}},\\
        e_A &= e_\Omega + c_2 e_R
\end{aligned}
\end{equation}
where $\eta$ is a small positive constant.
\subsection*{Proof}
Consider the following Lyapunov candidate:
\begin{equation}\label{eqn:lyapunov_p_hat}
\begin{aligned}
          \mathcal{V}_p =~& \frac{1}{2}e_\Omega^T {H_p}e_\Omega + k_R\Phi(R,R_d) + c_2 {e}_R \cdot {H_p}  {e}_\Omega + d_\psi ({h}_p \| {\hat{h}})
\end{aligned}
\end{equation}
Differentiating $\mathcal{V}_p$ along the solutions of the system and employing (\ref{eqn:control_torqueSMC_app}), we obtain
\begin{align*}
    \mathcal{\dot{V}}_p =& {e}_\Omega^T {H_p} {\dot{e}}_\Omega + k_R {e}_R \cdot {e}_\Omega + c_2 {\dot{e}}_R \cdot {H_p} {e}_\Omega \\
    &+ c_2 {e}_R \cdot {H_p} {\dot{e}}_\Omega + \dot{d_\psi }(\cdot) \\
     =& -k_\Omega {e}_\Omega ^T {e}_\Omega + {e}_\Omega ^T {Y}({\hat{h}}_p - {h}_p) \\
     & + e_\Omega^T(\Delta + \mu) +
      c_2 C(R_d^T,R){e}_\Omega \cdot {H_p} {e}_\Omega \\
      & - c_2 k_R {e}_R^T {e}_R + c_2 {e}_R \cdot {Y}({\hat{h}}_p - {h}_p)  + c_2 e_R^T(\Delta + \mu) \\
      & -c_2 k_\Omega {e}_R \cdot {e}_\Omega + \dot{d_\psi }(\cdot) \\
     =& -k_\Omega {e}_\Omega ^T {e}_\Omega +  c_2 C(R_d^T,R){e}_\Omega \cdot {H_p} {e}_\Omega - c_2 k_R {e}_R^T {e}_R \\
     &+ {e}_A \cdot {Y}({\hat{h}}_p - {h}_p)  + e_A \cdot (\Delta + \mu) \\
     & - c_2 k_\Omega {e}_R \cdot {e}_\Omega +  \dot{d_\psi }(\cdot)
\end{align*}
Since $\norm{C(R_d^T,R)} \leq \frac{1}{2}\textbf{tr}[G]$, we obtain
\begin{equation}\label{eqn:Vdot_robust}
    \begin{aligned}
    \mathcal{\dot{V}}_p =~& -k_\Omega {e}_\Omega ^T {e}_\Omega  - c_2 k_R {e}_R^T {e}_R + c_2 C(R_d^T,R){e}_\Omega \cdot {H_p} {e}_\Omega
    \\
     &- c_2 k_\Omega {e}_R \cdot {e}_\Omega \\
    \leq~& -\Big( k_\Omega - \frac{c_2}{\sqrt{2}} \Lambda_{p_{max}} \text{tr}[G]\Big ) \| {e}_\Omega \|^2 - c_2 k_R  \| {e}_R \|^2  \\
    &+ c_2 k_\Omega \| {e}_R \| \| {e}_\Omega\| + e_A \cdot (\Delta + \mu) 
    \end{aligned}
\end{equation}
The last term in above equation is bounded by
\begin{equation}
    \begin{aligned}
        e_A \cdot (\Delta + \mu ) & \leq \norm e_A \delta_R   - \Big(\delta_R - \frac{\eta}{ \norm{e_A}}\Big)\frac{e_A \cdot e_A}{\norm{e_A}} \\
        & \leq \eta
    \end{aligned}
\end{equation}
which implies $\dot{\mathcal{V}}_p$ is bounded by
\begin{equation}
    \begin{aligned}
    \mathcal{\dot{V}}_p \leq -z_1^T W_{31}^p z_1 + \eta
    \end{aligned}
\end{equation}
Hence if $\eta$ is adaptively chosen such that \\ $ \eta < z_1(t)^T W_{31}^p z_1(t)$, the attitude errors are asymptotically stable.
\begin{figure}[!t]
    \centering
    \subfloat[Tracking of angular velocity]{\includegraphics[width = 0.48\textwidth]{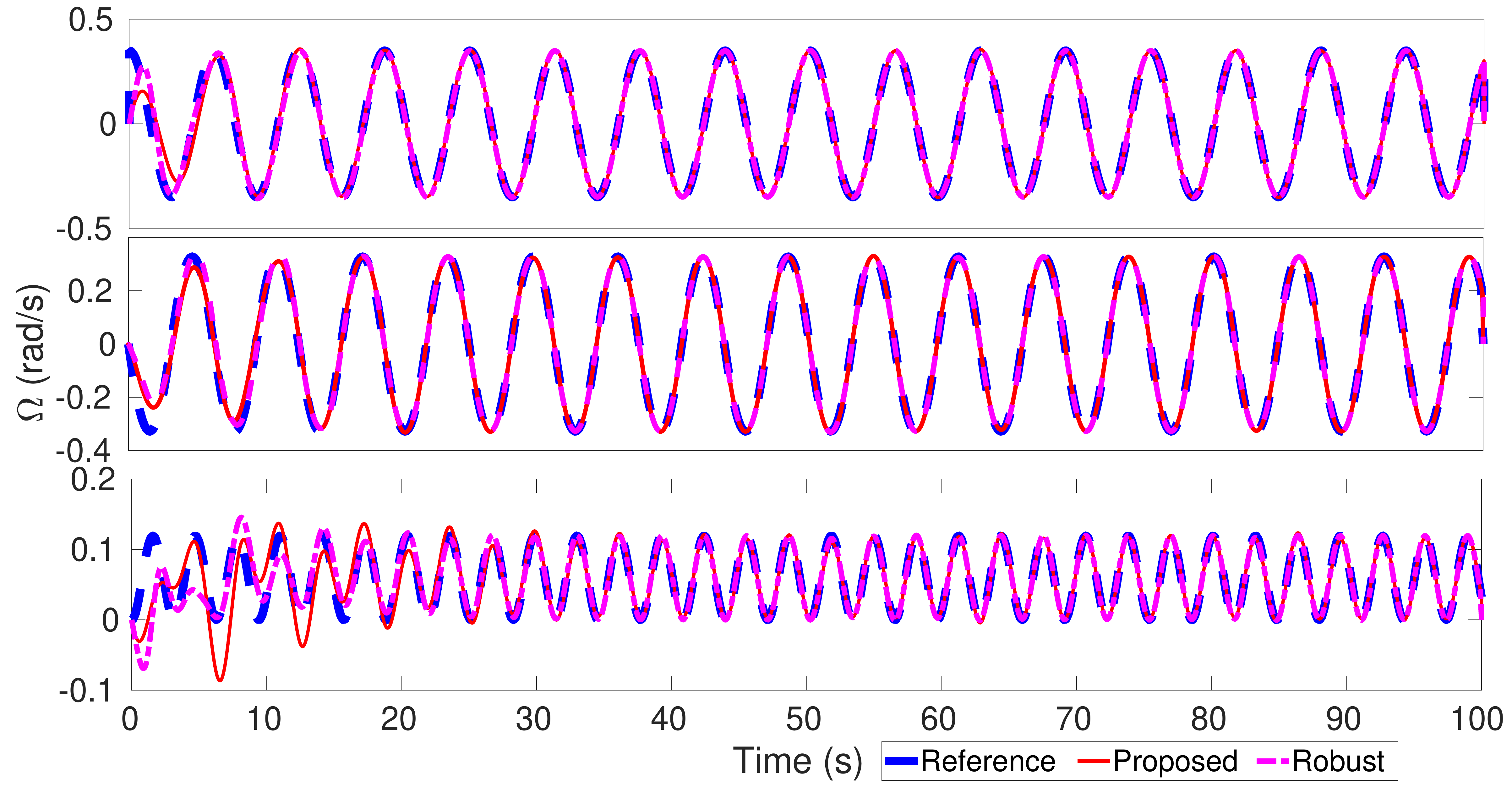}}\\
    \subfloat[Control effort ]{\includegraphics[trim = 0 2cm 0 0, clip,width = 0.48\textwidth]{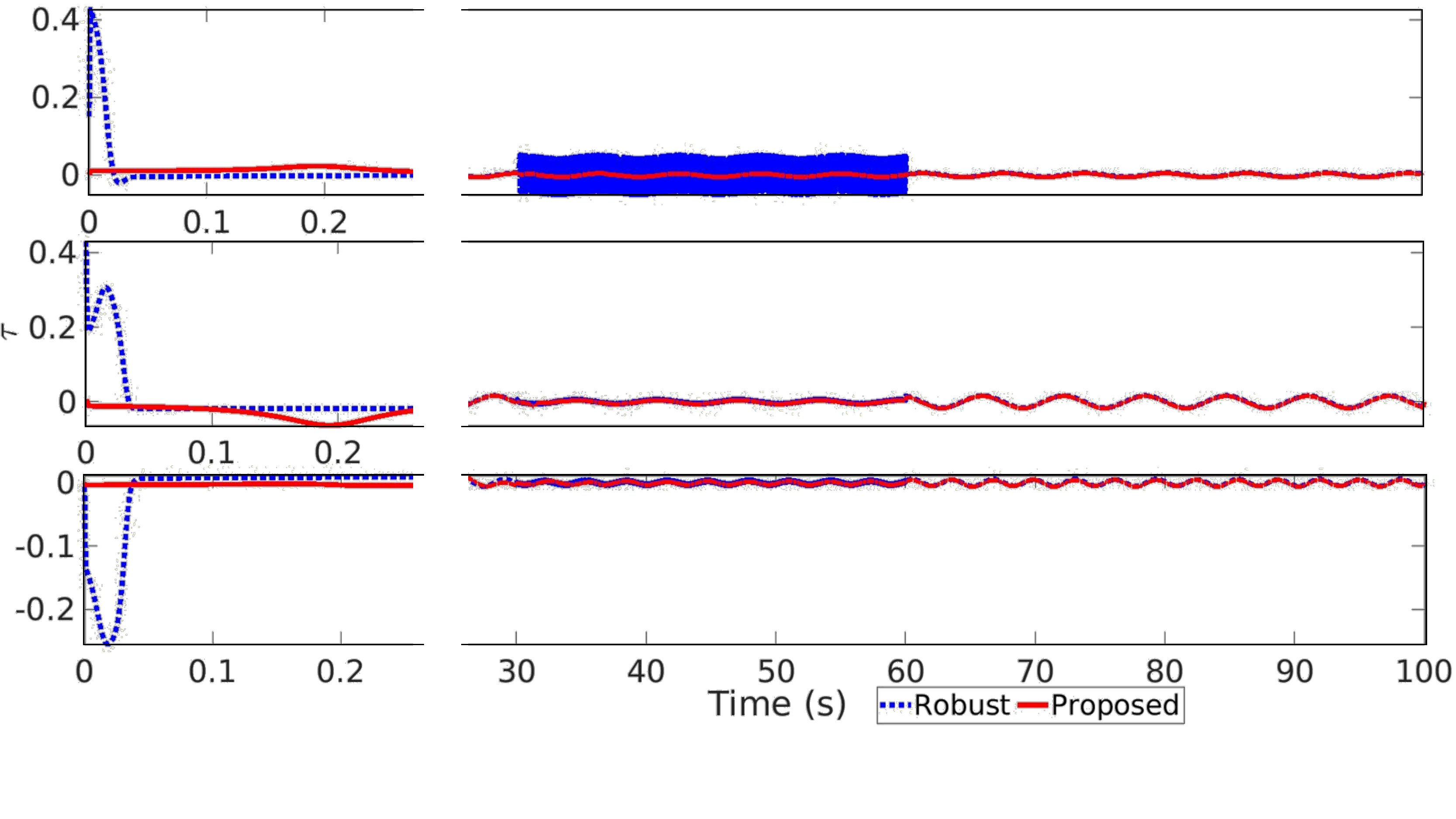}}
    \caption{Performance comparison between the proposed adaptive controller and a conventional robust  controller \cite{L+18} \textit{, without the presence of disturbances,} when the configuration switches from 1 to 2 at $t = 30$s and from 2 to 1 at $t = 60$s. The uncertainty bounds assumed for the robust controller are constant across the two subsystems and are too high for the subsystem 2. This leads to high control efforts and chattering. It also demonstrates higher control efforts initially from t = 0 to 1s. The adaptive controller has similar performance, however, with low control efforts through-out the entire flight. Please note that, we define control effort as the magnitude of the control torques computed by the controller.}
    \label{fig:robust_controller}
\end{figure}
Here, we choose to employ a bounded disturbance for the model in order to decrease the control efforts (we define control effort as the magnitude of the control torques and thrust, as computed by the controller) and minimise discontinuity. Other discontinuous disturbance rejection laws can also be employed of the form
\[ \mu = -k_\mu ~sign(e_A)\]
at the expense of conservative and discontinuous control efforts, where $k_\mu = \delta_R + \nu$ for a small positive constant $\nu$ and $sign(e_A)$ represents the sign of each element of the vector $e_A$.
\subsection{Results}
This section describes the various case scenarios simulated to validate the proposed controller for the switched system.
\subsubsection{Simulation Setup for Figure \ref{fig:adaptive_controller}} To validate Case 2 with no external disturbances, the inertia matrix (${H}_p$) for the $p^{th}$ configuration is assumed to have the structure ${H}_p = {H}_p^0 + \Delta {H}_p$ with the nominal inertia matrix, ${H}_p^0$, and uncertainty $\Delta {H}_p$. The values for ${H}_p^0$ are obtained from the formulations in our previous work \cite{PM+20} for two configurations 1 and 2 as
 corresponding to $l_1 = 0.2$ and $l_2 = 0.1$. The rest of the parameters are set to $
M = 1.4\text{kg}, k_R = 0.0424, k_\Omega = 0.0296
$.
The initial guess in the inertia parameters for the estimator are chosen as the nominal matrix ${H}_p^0$, more specifically:
\begin{equation}\label{eqn:nomInertia}
	\begin{aligned}
		\hat{H}_1(0) = & \begin{bmatrix}
			0.0023 &  -0.0006  &  0.0010\\
			-0.0006 &   0.0172   &      0\\
			0.0010   &      0 &   0.0181
		\end{bmatrix},\\
		\hat{H}_2(0) = & \begin{bmatrix}
			0.0014  &  -0.0001   &  0.0005\\
			-0.0001  &   0.0052   &       0\\
			0.0005    &      0    & 0.0053
		\end{bmatrix}
	\end{aligned}
\end{equation}
with the uncertainty set to:
\begin{equation*}
	\Delta {H}_1 = \Delta {H}_2 = \text{diag}[0.01,0.01,0.02]
\end{equation*}
\subsubsection{Comparison against Conventional Robust Controller}
Here, we present the comparison between the performances of the proposed adaptive controller in Section \ref{sec:caseB} and a conventional robust controller that accounts for parameter-varying uncertainty in inertia by assuming an upper bound.
For this case, we retain the nominal matrix from (\ref{eqn:nomInertia}):
\[ {H}_1^0 = \hat{H}_1(0), {H}_2^0 = \hat{H}_2(0) \]
However, the actual uncertainty in parameters for this case are chosen as 
\begin{equation*}
	\Delta {H}_1 = \text{diag}[0.2,0.2,0.4], \Delta {H}_2 = \text{diag}[0.1,0.1,0.2]
\end{equation*}
with the assumed bound as $\delta_R = 0.5$ for the robust controller.
The results for angular velocity tracking and the corresponding control efforts are seen in Fig. \ref{fig:robust_controller}(a)-(b). Please note, we define control effort as the magnitude of the control torques and thrust, as computed by the controller. From Fig. \ref{fig:robust_controller}, it is seen that the performance of the proposed adaptive controller is comparable to that of the robust controller, however with low control efforts. Further, the robust controller uncertainty assumed ($\delta_R = 0.5)$ is too high for the subsystem 2 which leads to chattering as shown in Fig. \ref{fig:robust_controller}(b). 
\begin{figure}[!t]
\centering
     \subfloat[Angular velocity tracking]{\includegraphics[width = 0.48\textwidth]{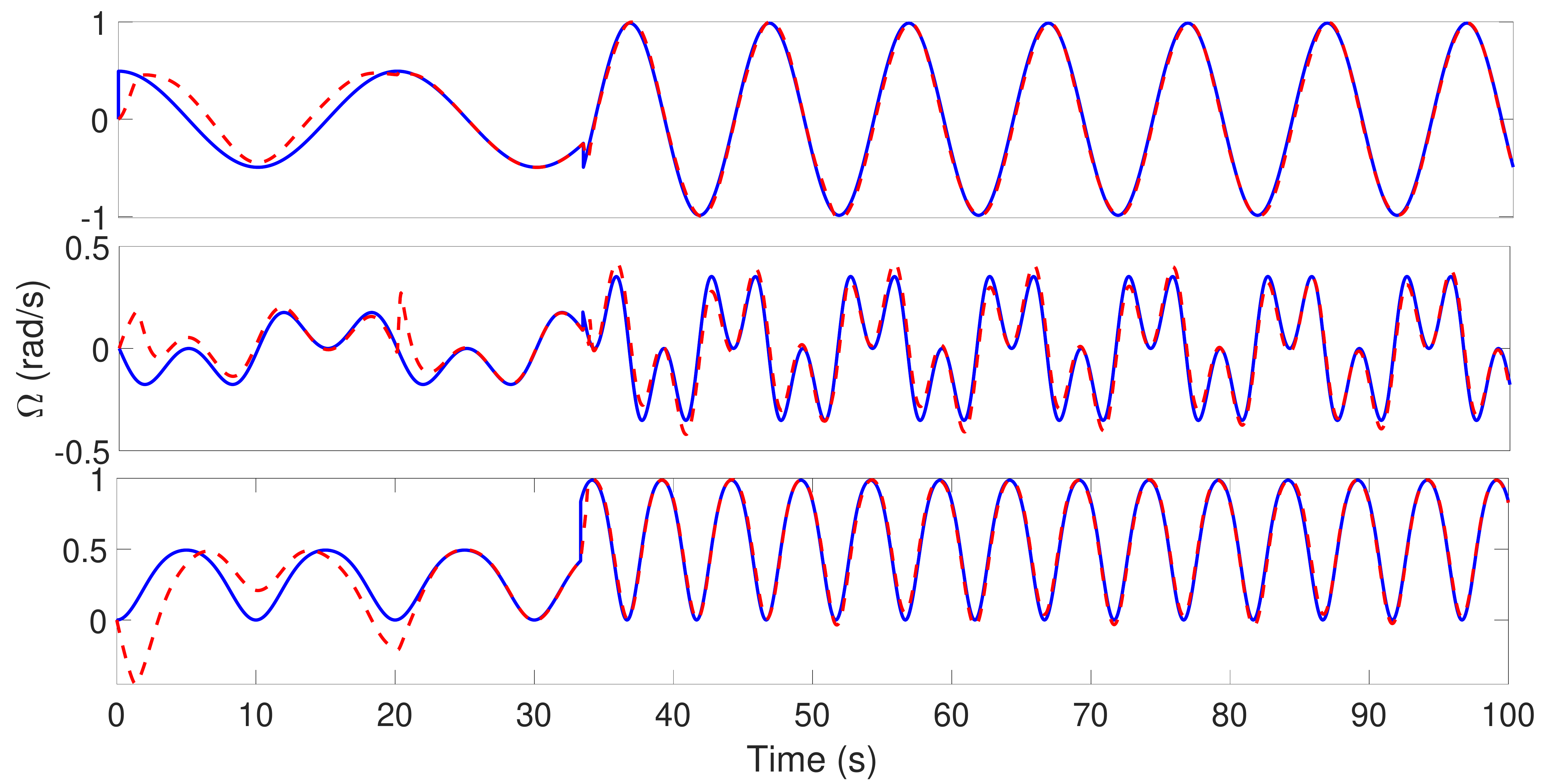}}\\
    \subfloat[Errors in ${e}_R$]{\includegraphics[width = 0.48\textwidth]{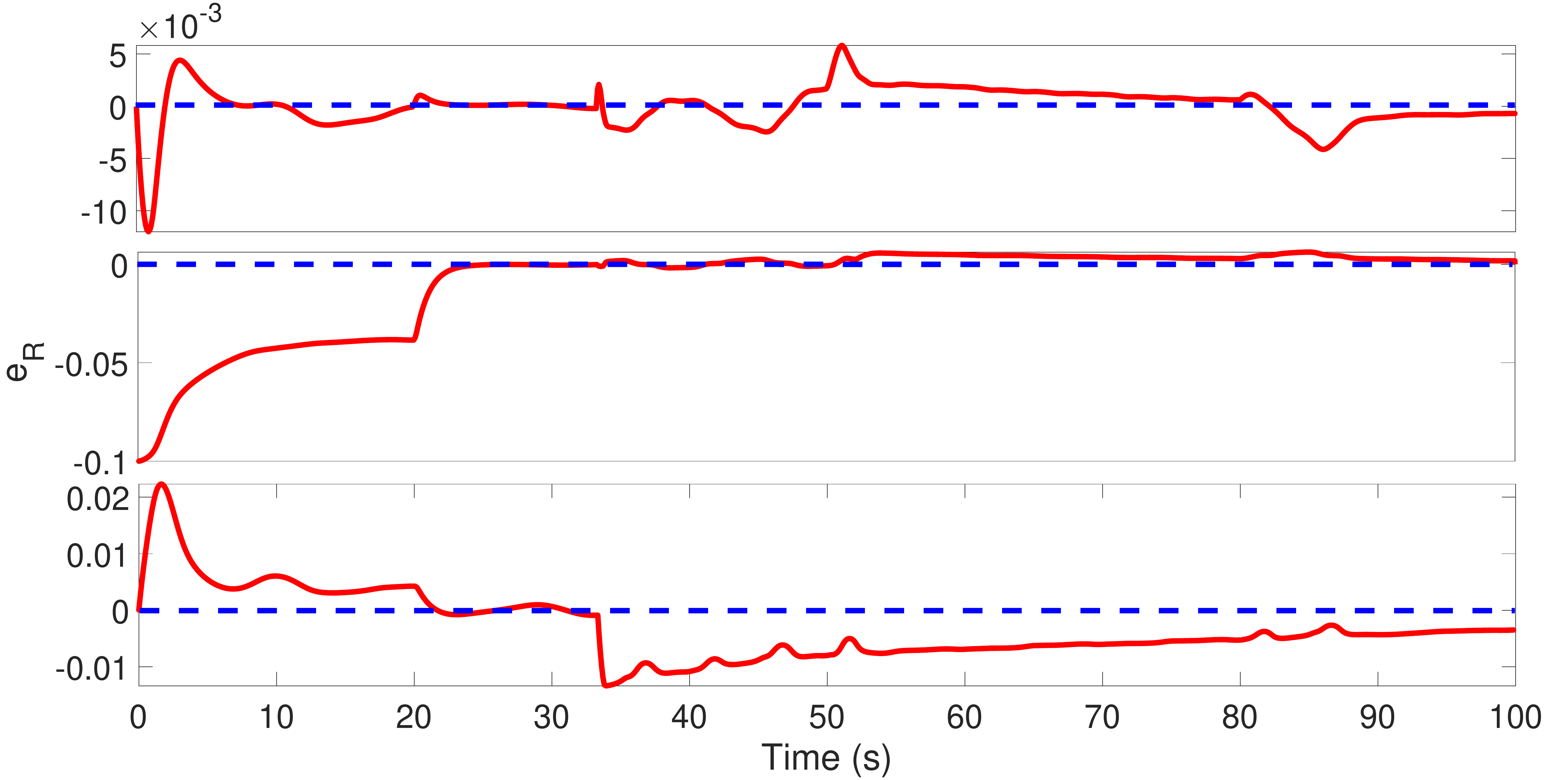}}
    \vspace{-0.1in}
    \\
     \subfloat[Inertia estimates (kg-m$^2$)]{\includegraphics[width = 0.48\textwidth]{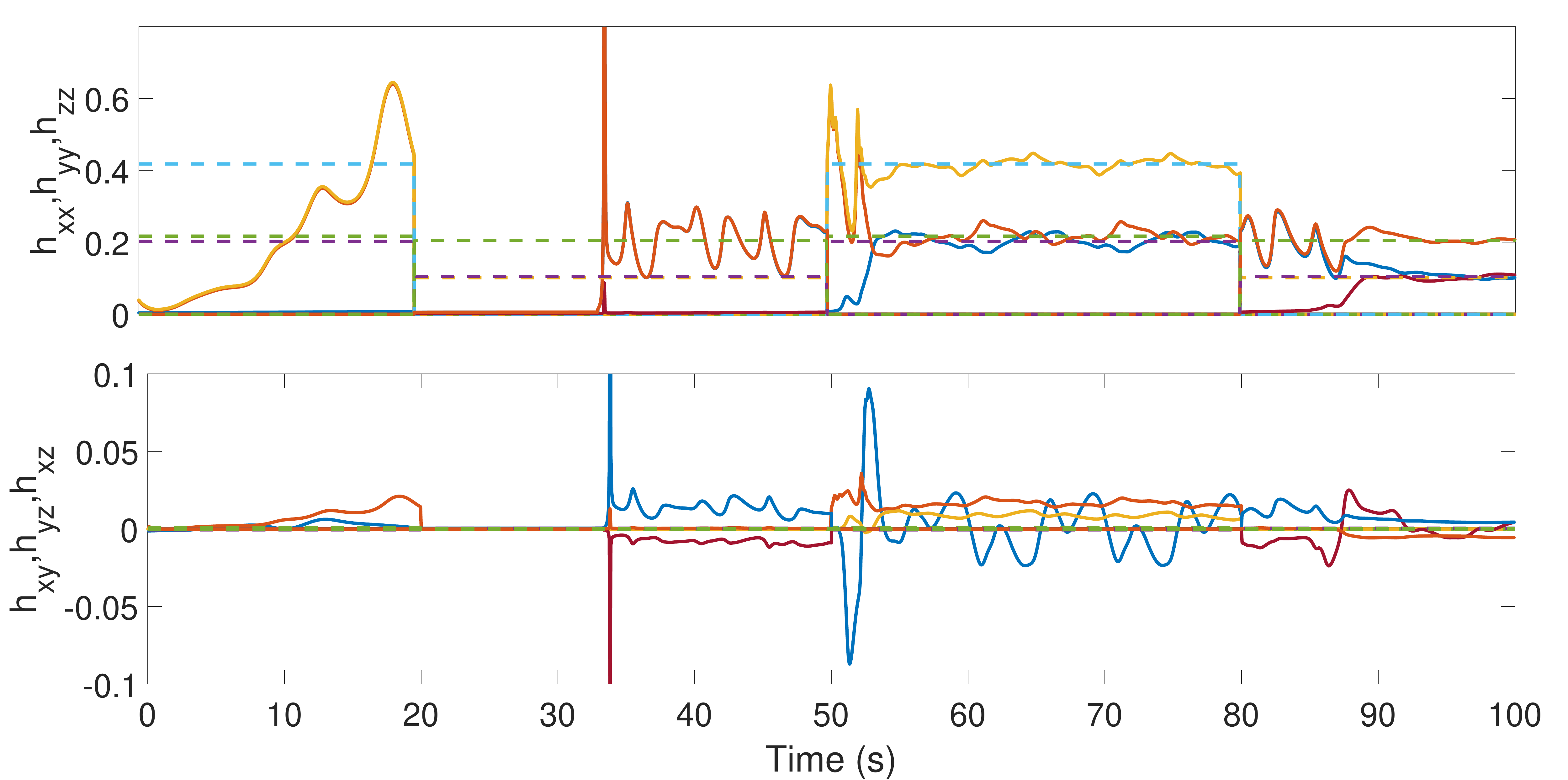}}\\
     \subfloat[Control effort]{\includegraphics[width = 0.48\textwidth]{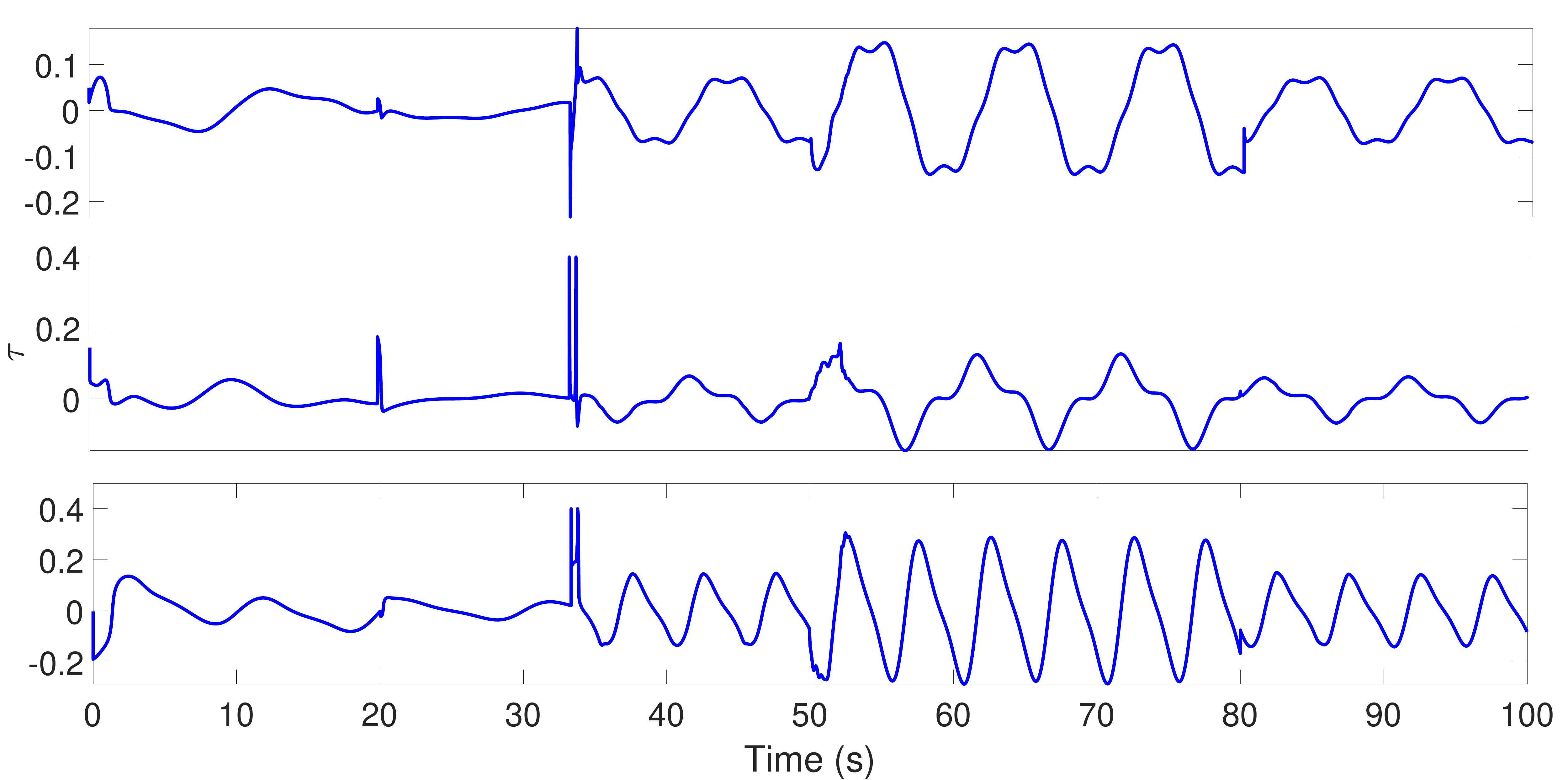}}
     \vspace{-0.1in}
    \caption{The performance of the proposed robust adaptive controller in the presence of unmodeled aerodynamic disturbances, assumed as bounded uncertainties. The tracking of the angular velocity and the attitude errors $e_R$ show that the zero equilibrium of the attitude tracking error is reached asymptotically. The inertia estimates also converge in this case.}
    \label{fig:rAd_results}
    \vspace{-0.3in}
\end{figure}
\subsubsection{Results for Proposed Controller in Sec. \ref{sec:caseC}}
Since the adaptive controllers can be unstable even for a slight disturbance, we simulate the case when there is disturbance added to the system with $\Delta (t) = 0.1[0~sin(t)~cos(t)]^T$. Remaining parameters are retained from Section I2 above. The proposed robust adaptive controller in (\ref{eqn:control_torqueSMC_app}), is simulated with low gains as above and low uncertainty bounds assumed at $\delta_R = 0.2$ and $\eta = 0.0003$. We also employ a varying frequency reference to ensure that the tracking errors are high after the quadrotor switches to a different subsystem to facilitate convergence of the inertia estimation errors. 

The tracking of the angular velocity and the attitude errors $e_R$ show that the zero equilibrium of the attitude tracking error is reached asymptotically with the proposed adaptive controller augmented with the robust term. The inertia estimates also converge to the true inertia values (although this is not guaranteed, as mentioned due to the lack of persistence of excitation). The logarithmic nature of the estimation error leads to a longer delay in the tracking of the inertia parameters for subsystem 2 as shown by the maroon solid curve in Figure \ref{fig:rAd_results}(c). In our future work, we would like to extend our analysis to improve the convergence rate and robustness in the presence of any matched input uncertainties. 

\begin{figure}[t]
\centering
     \subfloat[Angular velocity tracking]{\includegraphics[width = 0.48\textwidth]{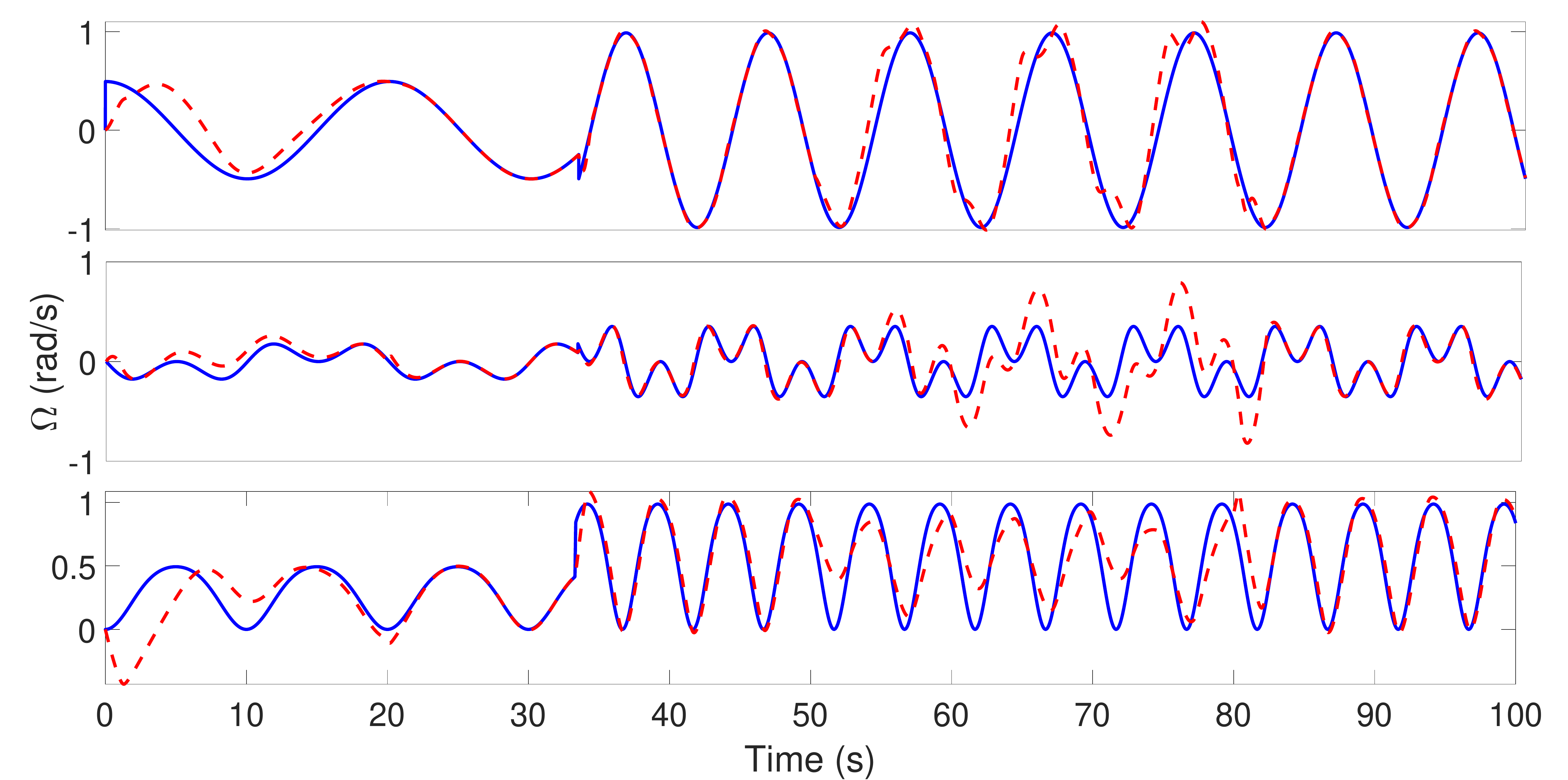}}\\
    \subfloat[Errors in ${e}_R$]{\includegraphics[width = 0.48\textwidth]{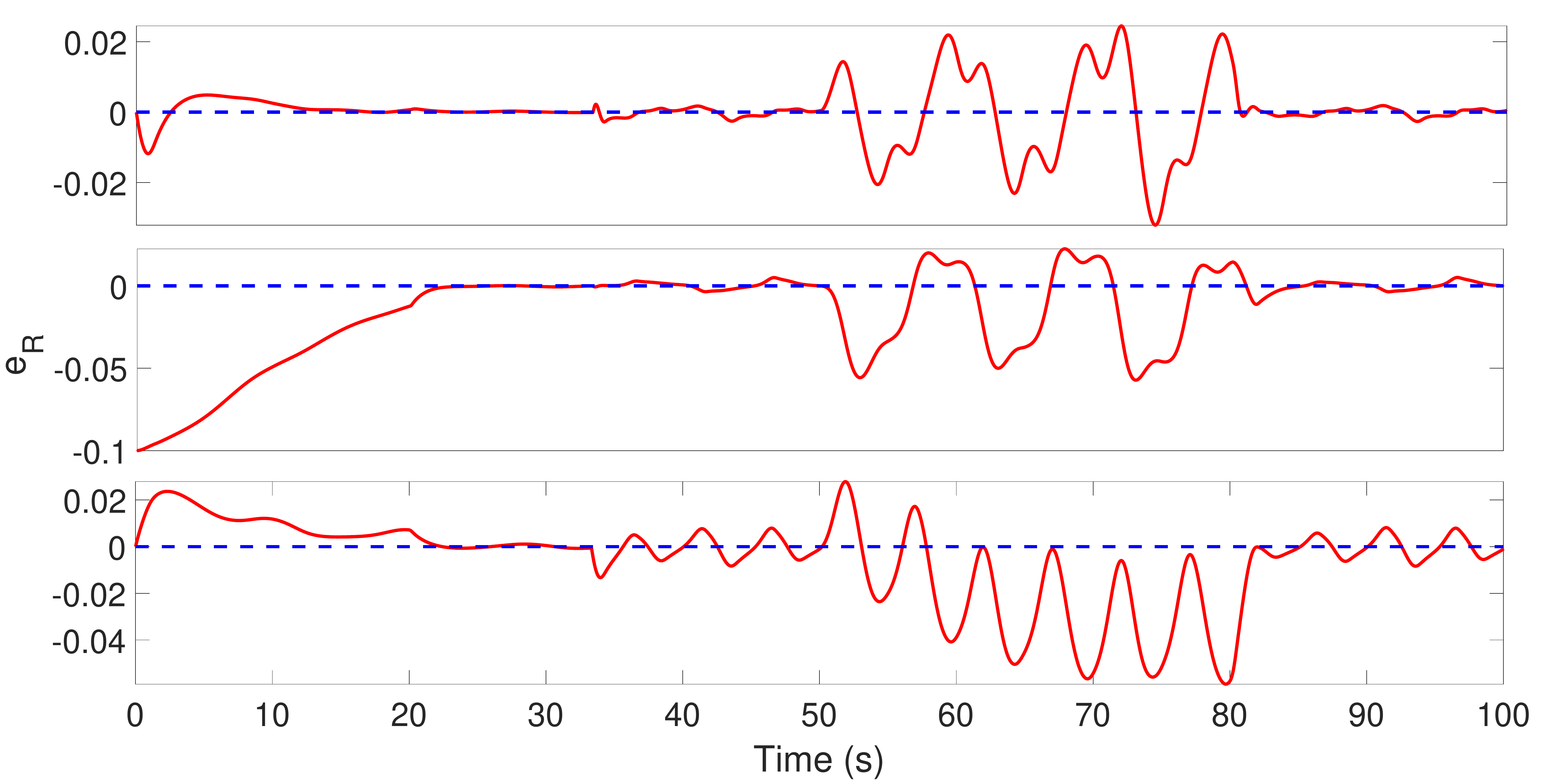}}
    \vspace{-0.1in}
    \\
     \subfloat[Control effort]{\includegraphics[width = 0.48\textwidth]{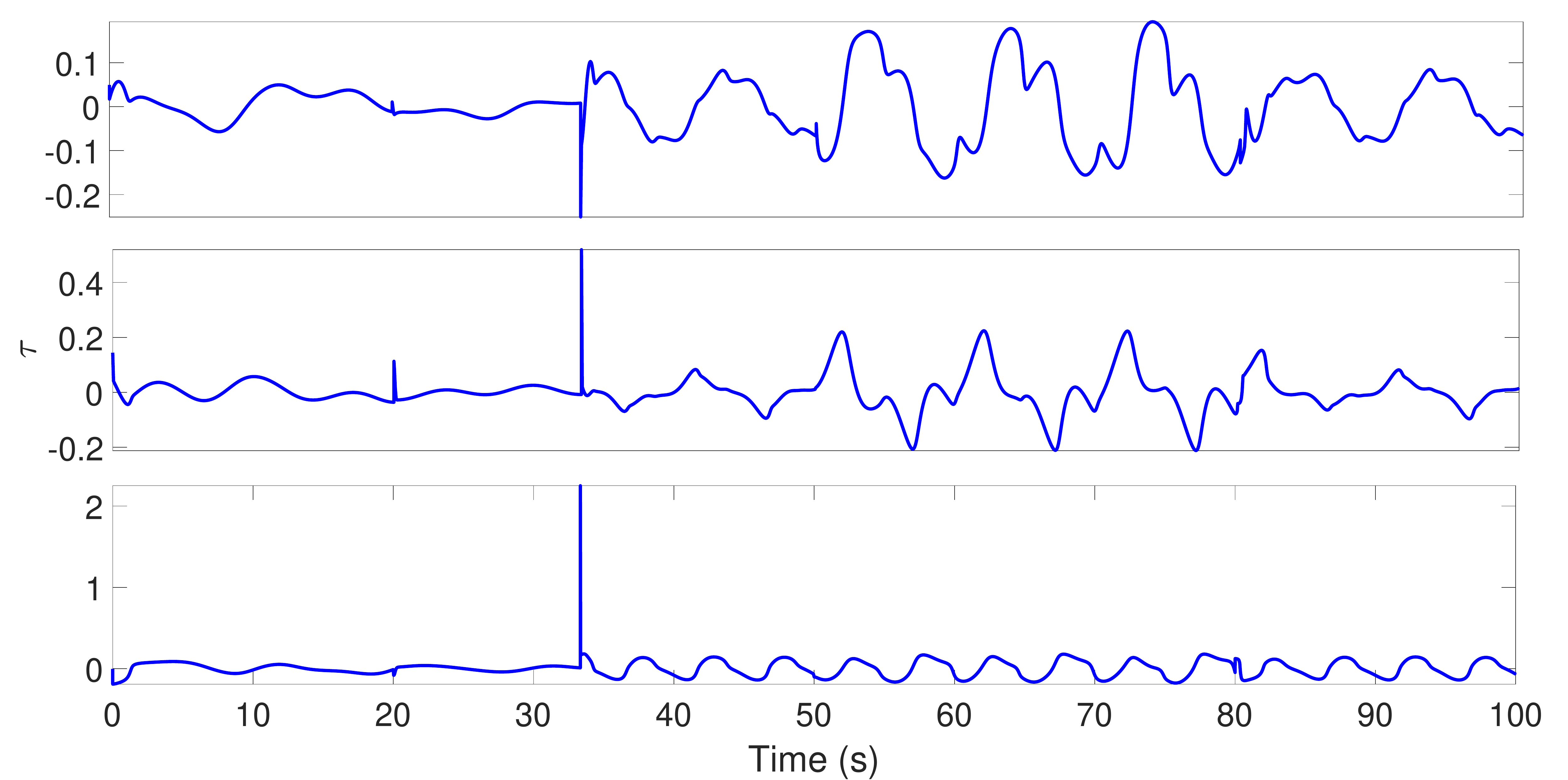}}
     \vspace{-0.1in}
    \caption{The performance of a conventional robust adaptive controller in the presence of unmodeled aerodynamic disturbances and model uncertainties, assumed as bounded uncertainties. The bound assumed is low for subsystem 1 and hence leads to a deterioration in the tracking performance.}
    \label{fig:rAd_comparison}
\end{figure}

Furthermore, from the Figs. \ref{fig:rAd_results} and \ref{fig:rAd_comparison}, it is seen that the proposed controller performs better against the conventional robust controller by leading to improved tracking performance as shown in Figure \ref{fig:rAd_results}(a)-(d), thereby validating \textit{Proposition 12}. The net bounded uncertainty ($\delta_R$) assumed is lower than the value coupled with the uncertainties in inertia for the conventional robust controller case. Hence the robust controller does not perform well in this case scenario. If the bounds assumed are too high, it may lead to chattering as shown in the Figure \ref{fig:robust_controller}(b).

\end{document}